\documentclass[conference]{IEEEtran}
\IEEEoverridecommandlockouts
\usepackage{cite}
\usepackage{amsmath,amssymb,amsfonts}
\usepackage{algorithmic}
\usepackage{graphicx}
\usepackage{textcomp}
\usepackage{xcolor}
\def\BibTeX{{\rm B\kern-.05em{\sc i\kern-.025em b}\kern-.08em
    T\kern-.1667em\lower.7ex\hbox{E}\kern-.125emX}}

\usepackage{booktabs}% http://ctan.org/pkg/booktabs

\usepackage{array}
\usepackage{calc}
\usepackage{algorithm}
\usepackage{tikz}
\usepackage{filecontents}
\usepackage{subcaption}
\usepackage{cleveref}
\usepackage[font=small,labelfont=bf]{caption}
\usepackage{authblk}
\usepackage{tikz}
\newcommand*\circled[1]{\tikz[baseline=(char.base)]{
            \node[shape=circle,draw,inner sep=0.5pt] (char) {#1};}}
            
\begin{document}

\title{RDMAbox : Optimizing RDMA for Memory Intensive Workload*\\
}


\author[1]{Juhyun Bae}
\author[1]{Ling Liu}
\author[1]{Yanzhao Wu}
\author[2]{Gong Su}
\author[2]{Arun Iyengar}
\affil[1]{Georgia Institute of Technology, Atlanta, GA30329, USA}
\affil[2]{IBM Thomas J. Watson Research, New York, USA}
\date{}                     %% if you don't need date to appear
\setcounter{Maxaffil}{0}
\renewcommand\Affilfont{\itshape\small}

\maketitle

\begin{abstract}
We present RDMAbox, a set of low level RDMA optimizations that provide better performance than previous approaches. The optimizations are packaged in easy-to-use kernel and user space libraries for applications and systems in data center. We demonstrate the flexibility and effectiveness of RDMAbox by implementing a kernel remote paging system and a user space file system using RDMAbox. RDMAbox employs two optimization techniques. First, we suggest RDMA request merging and chaining to further reduce the total number of I/O operations to the RDMA NIC. The I/O merge queue at the same time functions as a traffic regulator to enforce admission control and avoid overloading the NIC. Second, we propose Adaptive Polling to achieve higher efficiency of polling Work Completion than existing busy polling while maintaining the low CPU overhead of event trigger. Our implementation of a remote paging system with RDMAbox outperforms existing representative solutions with up to 4$\times$ throughput improvement and up to 83\% decrease in average tail latency in bigdata workloads, and up to 83\% reduction in completion time in machine learning workloads. Our implementation of a user space file system based on RDMAbox achieves up to 5.9$\times$ higher throughput over existing representative solutions.
\end{abstract}

%\begin{IEEEkeywords}
%component, formatting, style, styling, insert
%\end{IEEEkeywords}

\section{Introduction}
The tremendous growth in the amount of data that must be processed has resulted in dramatic increase in the demand for memory and storage by data intensive applications such as relational database, key-value store, and machine learning model training, etc. A cluster of machines is often required to satisfy the demand of such applications. RDMA is an interconnect technology that provides high throughput and low latency access to remote memory without CPU intervention. Recently, RDMA has been getting more attention and applied to data center applications and systems such as remote memory disaggregation systems~\cite{Infiniswap,nbdx, valet,LITE}, remote file systems~\cite{octopus,orion,glusterfs} and key-value stores~\cite{Storm,HERD14,FARM,FASST,DRTMH,PILAF}. RDMA is also well utilized in kernel space in remote memory disaggregation~\cite{Infiniswap,nbdx,valet,LITE} and kernel based distributed file systems~\cite{octopus}. However, building such systems with a native RDMA library requires considerable low level knowledge of the RDMA NIC.

Several RDMA optimizations have been introduced through research efforts in many different systems including above mentioned systems\cite{octopus,Storm,HERD14,FARM,FASST,DRTMH,PILAF,Infiniswap,nbdx,HERD16,philip2009,LITE}. However, those techniques are not fully optimized at a networking stack, and there is considerable room for improvement. Moreover, those techniques are applied in different systems and are not compared with each other. This makes it hard for application and system developers to understand which optimization technique is suitable for which system design.

% contribution 1 : RDMA optimizations
To further improve RDMA performance optimization, we focus on two main problems. First, the zero copy capability of RDMA offloading is a double edged sword. While it makes initiation of I/O operations faster, it can also overwhelm the limited onboard resources in the NIC and lead to inefficient usage of PCIe bus between the CPU and NIC (\cref{nicbottleneck}). Existing approaches such as doorbell batching discussed in recent research~\cite{nbdx, HERD16} chains multiple requests together and uses only one memory mapped I/O (MMIO) for the first request while the remaining requests are transferred via DMA. However, it does not reduce the total number of RDMA operations to the NIC(Table.\ref{rdmaio} in~\cref{experiments}). RDMAbox introduces merging on RDMA Memory Region(MR) and chaining that further improves batching efficiency over existing doorbell batch technique(\cref{messagebatching}). Merging and chaining opportunistically look for multiple adjacent requests that use contiguous memory addresses in the destination and merges them into a single request. Requests that have different memory addresses in the destination can be further chained if the destination node is same. Therefore, in addition to reducing the number of I/Os, merging and chaining also reduce the handshake between CPU and NIC.

Even with batching, the NIC's limited onboard resources can still be overwhelmed since the RDMA architecture lacks admission control (\cref{nicbottleneck}). Recent network congestion control solutions such as Timely \cite{TIMELY} can detect NIC overload by including delay in the NIC in the RTT calculation. But this incurs measuring overhead, and its expensive floating point calculation is not feasible in kernel space. RDMAbox implements a simple traffic regulator for admission control of RDMA I/O to the NIC by utilizing the I/O merge queue.(\ref{pacerdesign}). The traffic regulator stops I/O flow when the merge queue is filled to a configurable amount to prevent the NIC from being overwhelmed. We show that the simple yet effective traffic regulator provides 30\% higher throughput than the case without admission control under heavy RDMA I/O load.

Second, few research efforts have been made on analysis of Work Completion(WC) handling mechanisms in RDMA systems. We first reveal limitations of existing approaches in terms of CPU usage, parallelism and scalability (\cref{limitWC}). We then propose a new polling scheme called Adaptive Polling (\cref{adaptivepolling}). Adaptive Polling is triggered by completion events instead of running all the time, so it has lower CPU overhead than busy polling. Once triggered, Adaptive Polling will continue to poll the Completion Queue (CQ) up to a configured number of WCs or until no WC is left in the CQ. This is similar to Linux NAPI~\cite{NAPI,Accelio} or hybrid polling approach~\cite{X-RDMA}. However, these approaches return to event mode immediately when they fail to poll even though there is a pack of requests arriving with a short interval. Then, interrupt handling leads to suboptimal performance(Fig.~\ref{microrst}). When busy polling fails to poll, Adaptive polling iterates to poll more with pre-defined value(we call it a hook) to catch incoming events that arrives with a short interval. Then, Adaptive Polling stays busy polling to process them quickly and avoid unnecessary interrupt handling.

%contribution 2 : node level abstraction as well as network level abstraction, detailed analysis for decision making
 We demonstrate the flexibility and effectiveness of RDMAbox by implementing a remote paging system as a kernel space example and a user space distributed file system using RDMAbox. We also conduct extensive performance comparisons and analysis to help application and system developers better understand various tradeoffs and make better system design decisions (\cref{experiments}). Our experiments show that in both kernel remote paging and userspace file system cases, RDMAbox based implementations significantly outperform their respective existing systems(\cref{evaluation}). In particular, comparing to implementation with Accelio~\cite{Accelio} that is an RDMA I/O acceleration library, remote paging system using kernel-space-RDMAbox improves throughput by up to 4× and reduces average latency by up to 83\% and 99th tail latency by up to 98\%, respectively in bigdata workloads, and reduces completion time by up to 83\% in machine learning workloads. Our FUSE-based file system using user-space-RDMAbox achieves up to 3.7$\times$ higher write throughput and up to 5.9$\times$ higher read throughput over existing best practices.
 
The main contributions of this paper include:
\begin{itemize}
\item low level RDMA optimizations that outperform previous solutions, packaged in easy-to-use APIs for kernel and user space;
\item extensive performance test comparison with previous solutions and detailed analysis to help better understand various tradeoffs and make system design decisions;
\item demonstration of the flexibility and effectiveness of RDMAbox by implementing and providing a kernel remote paging system and a user space file system.
\end{itemize}

The rest of the paper proceeds with background, problem statement(\cref{softwarechallenge}), RDMAbox optimizations (\cref{designoverview}) and performance impact and detailed analysis of RDMABox optimizations with real world workload from applications (\cref{experiments}), followed by evaluation with various data center application and workload patterns (\cref{evaluation}) and conclusion (\cref{conclusion}).

%-------------------------------------------------------------------------------
\section{Background}
\label{background}

\textbf{Physical address.}
FaRM\cite{FARM} reported PTE cache miss and performance decrease as registered MR increases. Suggested solution was to use physical address to avoid PTE cache miss. LITE\cite{LITE} also utilizes physical address for MR by implementing RDMA abstraction in kernel space. Recent research effort Storm\cite{Storm} suggested CMA(Contiguous Memory Allocation) in user space.

\textbf{Doorbell batch.}
Doorbell batch is utilized in many research efforts\cite{HERD16,DRTMH,FASST} and is well-known batching technique that RDMA NIC provides. It helps to reduce the total bandwidth consumption on PCIe by replacing MMIO with DMA read~\cite{HERD16} because MMIO uses more bandwidth than DMA. We discuss pros and cons of doorbell batch and propose merging and chaining to further optimize the performance of RDMA NIC for various purposes(\cref{messagebatching}).

\textbf{Memcpy vs MR registration.}
Frey et al.(2009)\cite{philip2009} pointed out that MR registration for small MR($<$256KB in their report) has a significant overhead compared to memcpy to pre-allocated and registered MR. However, this is only correct in user space with virtual address. We reveal different results with physical address in kernel space and provide this as an option in our design(\cref{messagebatching}).

\textbf{Event Batch Polling}\cite{nbdx} is extended from event-triggered mode. It is similar to Linux NAPI~\cite{NAPI}, which uses finite number of budget for batched processing. When event is triggered, it polls up to N times per event unless it is failed and it can get K WC items, where 0$\leqslant$K$\leqslant$N. It polls in interrupt context same as vanilla Event-triggered mode does, but it can reduce the number of interrupts from K to 1 to process K WCs. In this way, it improves performance compared to Event-triggered mode. After processing K WCs, Event batch goes back to Event-triggered mode even if there are other WCs arrived a bit late in CQ. 

\textbf{Hybrid(event + busy) Polling}\cite{X-RDMA} switches between event mode and busy polling. It stays in busy polling and immediately goes back to event mode when there is no event in the queue. 

\textbf{One Shared CQ and busy polling}\cite{LITE} are used as an extension of Busy polling case. Since Busy polling approach has too much CPU overhead when increasing the remote connection, it only uses one shared CQ(SCQ in short for the rest of the paper) and one busy polling on the system. It can reduce CPU overhead compared to N Busy polling threads.  Since it relies on one serialized shared CQ and busy polling thread, it reduces parallelism. We leave detailed analysis and experiments about parallelism and scalability to \cref{experiments}.

%-------------------------------------------------------------------------------
\section{Software Challenges}
\label{softwarechallenge}

%---------------------------------------------
\subsection{Inefficiency of single I/O} % problem 1: batch latency , problem 2: # of RDMA I/O
\label{nicbottleneck}
%With asynchronous nature of RDMA, applications are able to maximize cpu efficiency by posting multiple parallel I/O. However, increasing parallel I/O can cause bottleneck in NIC due to limited resource such as WQE cache~\cite{FARM,HERD16} and leads to inefficient use of PCIe bandwidth. 

%When a WR is posted, CPU writes WQE(converted from WR) to NIC with MMIO. To post N WRs, CPU needs to writes MMIO N times on NIC. Many single I/O postings lead to inefficiency in use of PCIe bandwidth between NIC and CPU. Another problem is that, due to limited resource in NIC, such as WQE cache and Memory Protection Table(MPT), which stores permission information of each MR, many parallel single I/O posting likely causes NIC bottleneck(Figure~\ref{nic_bottleneck}). 

%To figure out performance impact of many parallel single I/O on NIC, we build remote memory system with Linux block device and RDMA. This virtual block device is connected to remote nodes in the cluster through RDMA. One can mount this virtual block device and provide remote memory with POSIX file interface. 

\begin{figure}[ht!]
\begin{subfigure}{0.23\textwidth}
 \centering
 \includegraphics[width=1\linewidth]{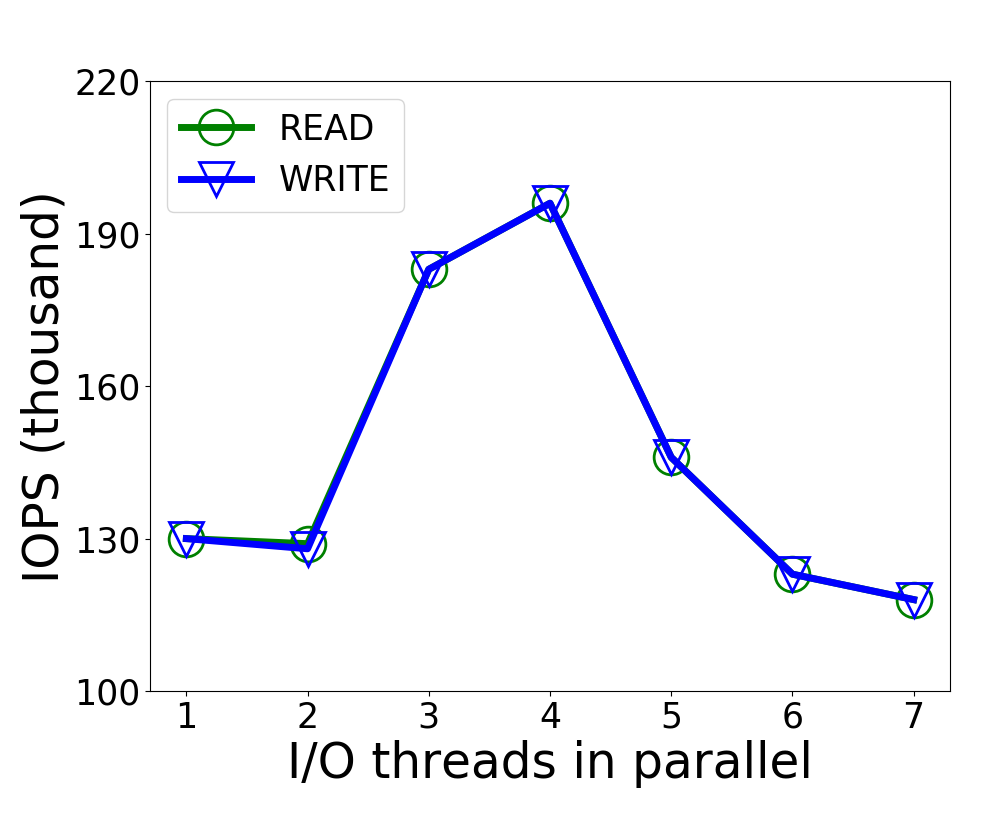}
\caption{IOPS}\label{nic_bottleneck:1a}
\end{subfigure}
\begin{subfigure}{0.23\textwidth}
 \centering
 \includegraphics[width=1\linewidth]{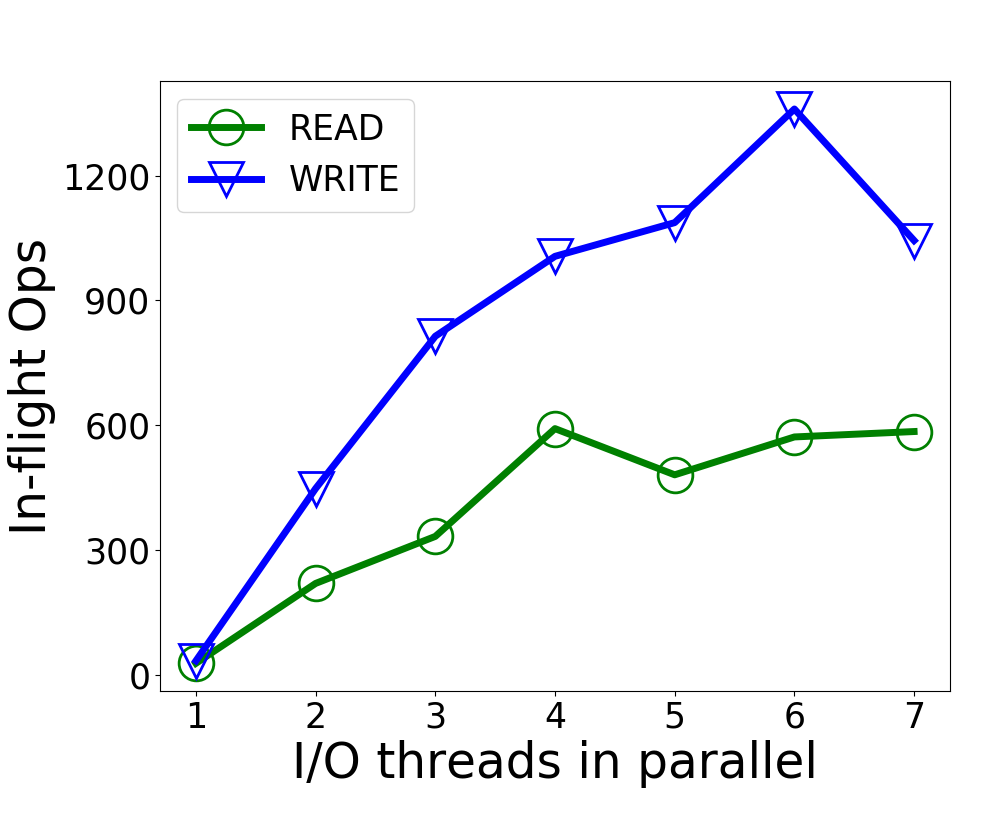}
\caption{RDMA in-flight ops}\label{nic_bottleneck:1b}
\end{subfigure}
\begin{subfigure}{0.23\textwidth}\vspace{11pt}
 \centering
 \includegraphics[width=1\linewidth]{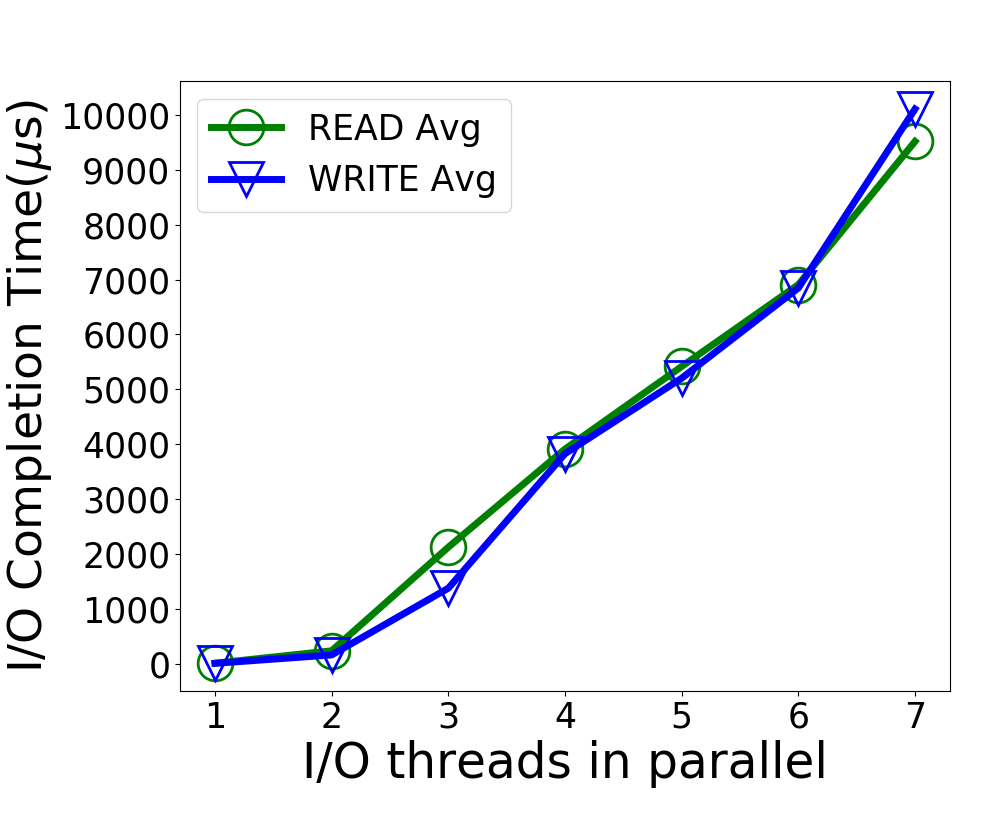}
\caption{RDMA I/O Completion time}\label{nic_bottleneck:1c}
\end{subfigure}\hspace{6pt}
\begin{subfigure}{0.23\textwidth}\vspace{11pt}
 \centering
 \includegraphics[width=0.84\linewidth]{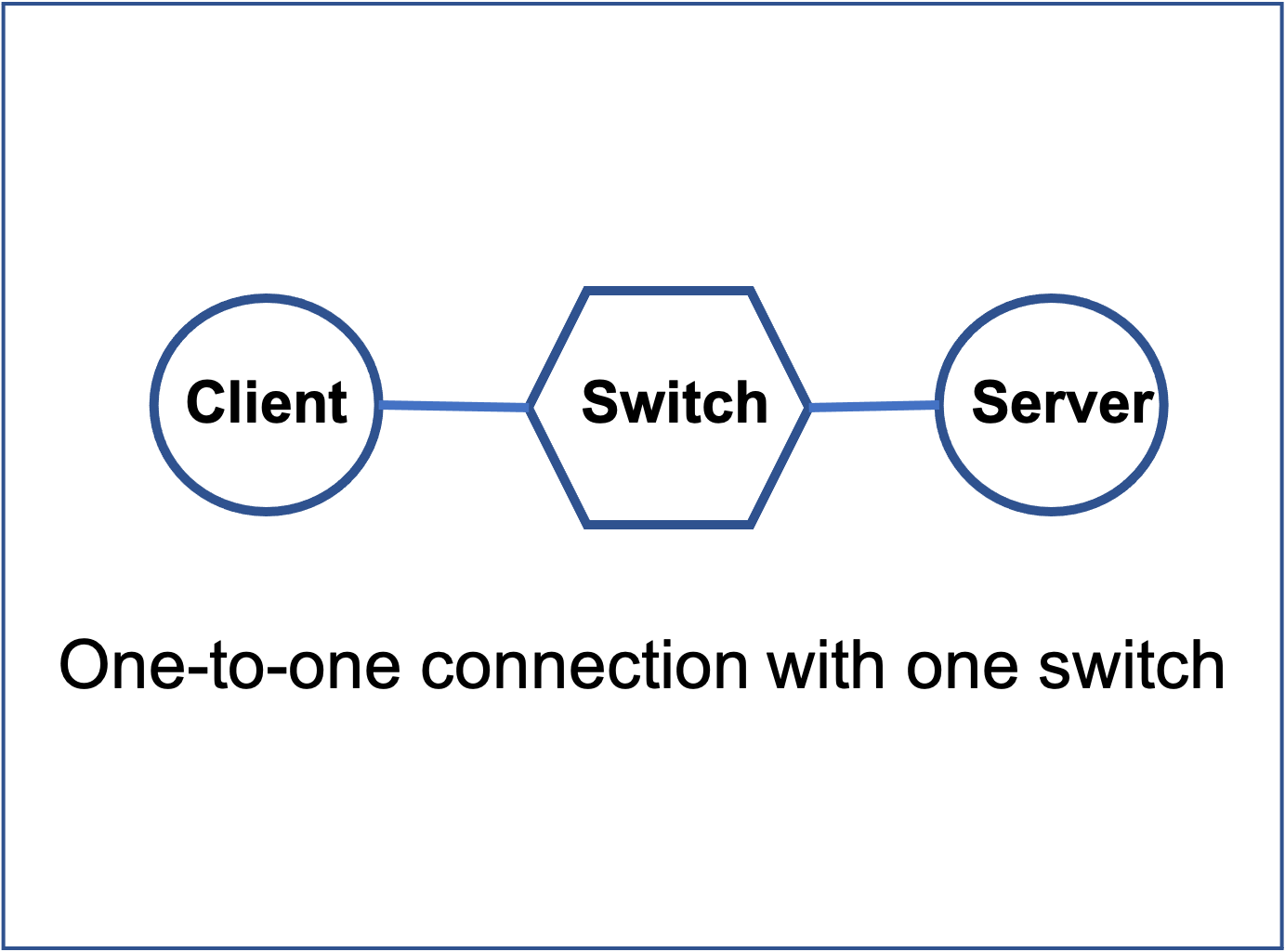}\vspace{11pt}
\caption{Experiment Setup}\label{nic_bottleneck:1d}
\end{subfigure}
\bigskip
\caption{IOPS starts to decrease at some point when workload increases(Fig.\ref{nic_bottleneck:1a}) as we see in the number of in-flight operations in RDMA stack keeps increasing(Fig.\ref{nic_bottleneck:1b}) and I/O completion time(Fig.\ref{nic_bottleneck:1c}) gets delayed more and more. This shows that NIC becomes bottleneck.}
\label{nic_bottleneck}
\end{figure}

We issue RDMA write operations to measure IOPS by varying the number of threads. At first, IOPS increases when the number of threads increase. At one point, IOPS starts to drop by increasing threads(Fig.\ref{nic_bottleneck:1a}). It shows that posting many parallel single I/Os can increase performance at first but it also can cause NIC bottleneck due to too many I/Os beyond NIC's capacity. Merging I/Os across data request threads can reduce the total number of I/O more and improves performance, but enforcing cross cpu/thread merging has a significant overhead. For instance, this is why Linux block layer does not provide cross-cpu I/O merging in the block layer~\cite{linuxcoalescing}. The benefit of parallel processing will be offeset by latency from merge-checking and batching. Although we give an example of kernel space case, applications who directly use RDMA library in user space also face the same I/O thrashing issue on NIC due to many parallel I/Os.

Handshake cost and limited resource of NIC are highly related to RDMA I/O performance. For each single I/O, it requires handshake between CPU and NIC such as doorbell memory write to put the request into NIC and DMA fetching. This handshake per single I/O becomes inefficient especially when the workload is very high and intense. NIC uses SRAM to cache many internal information such as connection state(QP), Work Requests(WQE) and address information(MPT). However, it is even limited in recent version of NIC, which causes more DMAs to fetch those information from main memory again\cite{erpc}. Vendors have not put more memory in NICs, because of cost, power overheads and market factors\cite{erpc}. Cache miss in NIC is fetched by DMA from main memory through PCIe bus, which makes performance suboptimal.

\vspace{-6pt}
%---------------------------------------------

\bigskip
\textbf{Lack of Admission Control for mitigating NIC bottleneck.}
Due to lack of proper management of I/O thrashing on NIC, it creates negative factors such as WQE cache miss to cause NIC bottleneck. This even can happen when network is not congested. 
In the experiment in Fig.\ref{nic_bottleneck}, we set only one client and one server node connected by a single switch, indicating no network congestion. When IOPS reaches peak point with 4 I/O threads, both RDMA I/O completion time and the number of in-flight ops are still increasing. This shows NIC becomes bottleneck and it takes longer time to process RDMA I/O requests with more parallel I/Os from user side. Lack of RDMA I/O level admission control can easily make RDMA I/O inefficient. Although Timely~\cite{TIMELY} can detect NIC bottleneck unlike other network congestion control, it is only for user space but not for kernel space due to expensive floating point operations for gradient based rate calculation. Moreover, operations are simply blocked while pacing the traffic, wasting potential chance(\cref{experiments}).

\vspace{-6pt}
\begin{figure}[!htb]
\begin{subfigure}{0.23\textwidth}
 \centering
 \includegraphics[width=1.1\linewidth]{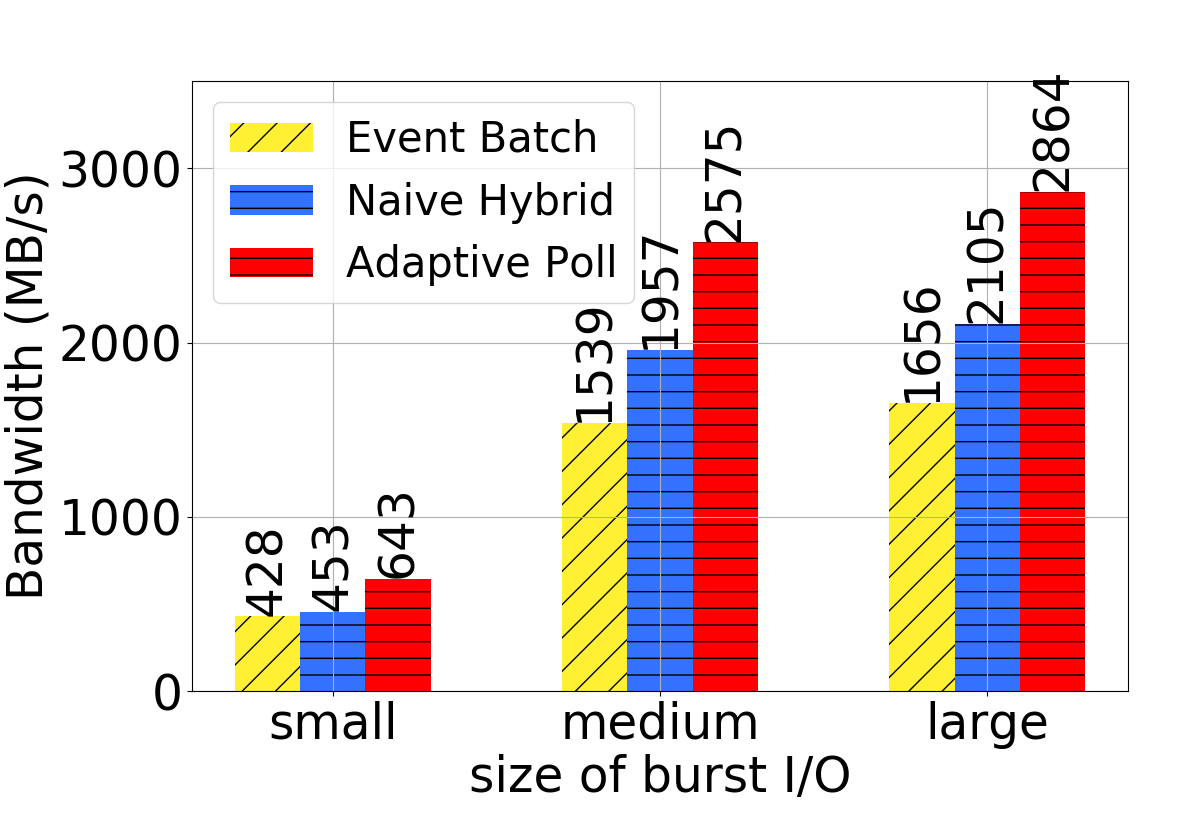}
  \caption{Bandwidth} \label{microrst:1a}
\end{subfigure}\hfill
\begin{subfigure}{0.23\textwidth}
 \centering
 \includegraphics[width=1.1\linewidth]{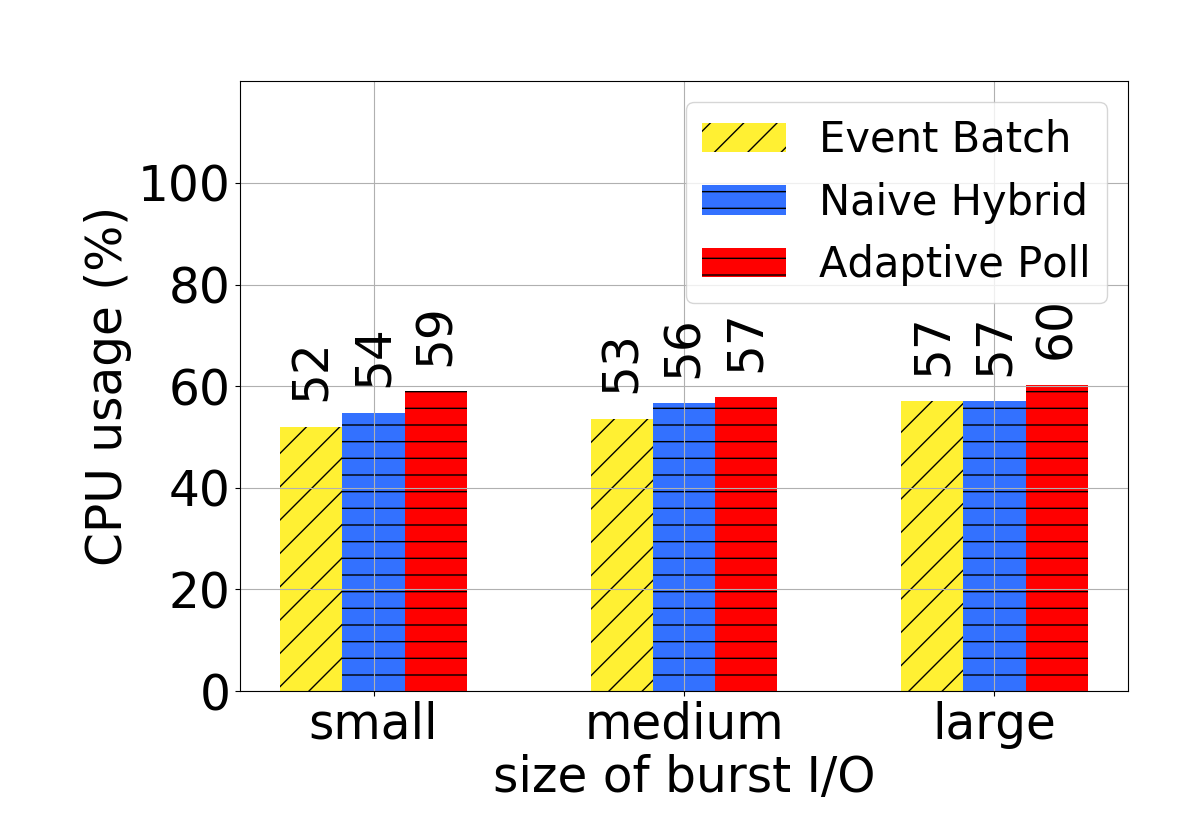}
 \caption{CPU usage} \label{microrst:1b}
\end{subfigure}\hfill
 \vfill
 \bigskip
 \begin{subfigure}{0.23\textwidth}
 \centering
 \includegraphics[width=1.1\linewidth]{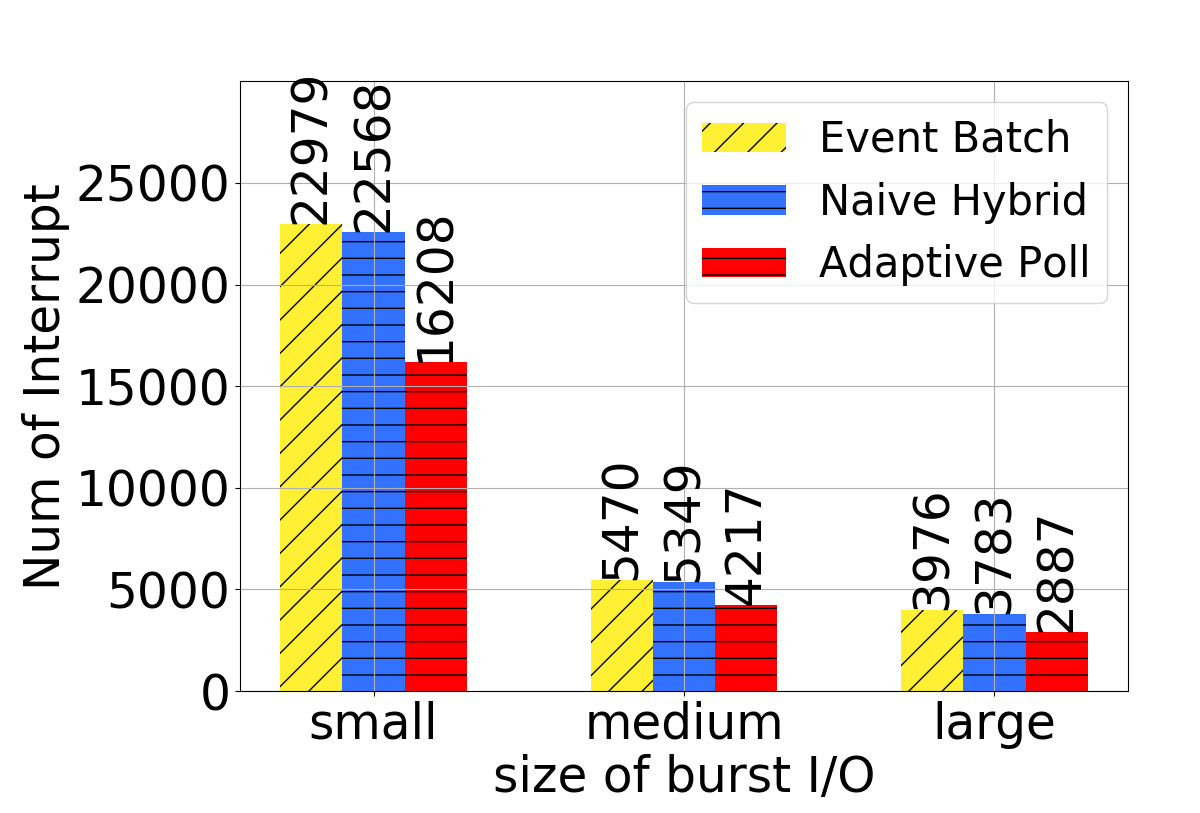}
 \caption{Interrupt} \label{microrst:1c}
\end{subfigure}\hfill
\begin{subfigure}{0.23\textwidth}
 \centering
 \includegraphics[width=1.1\linewidth]{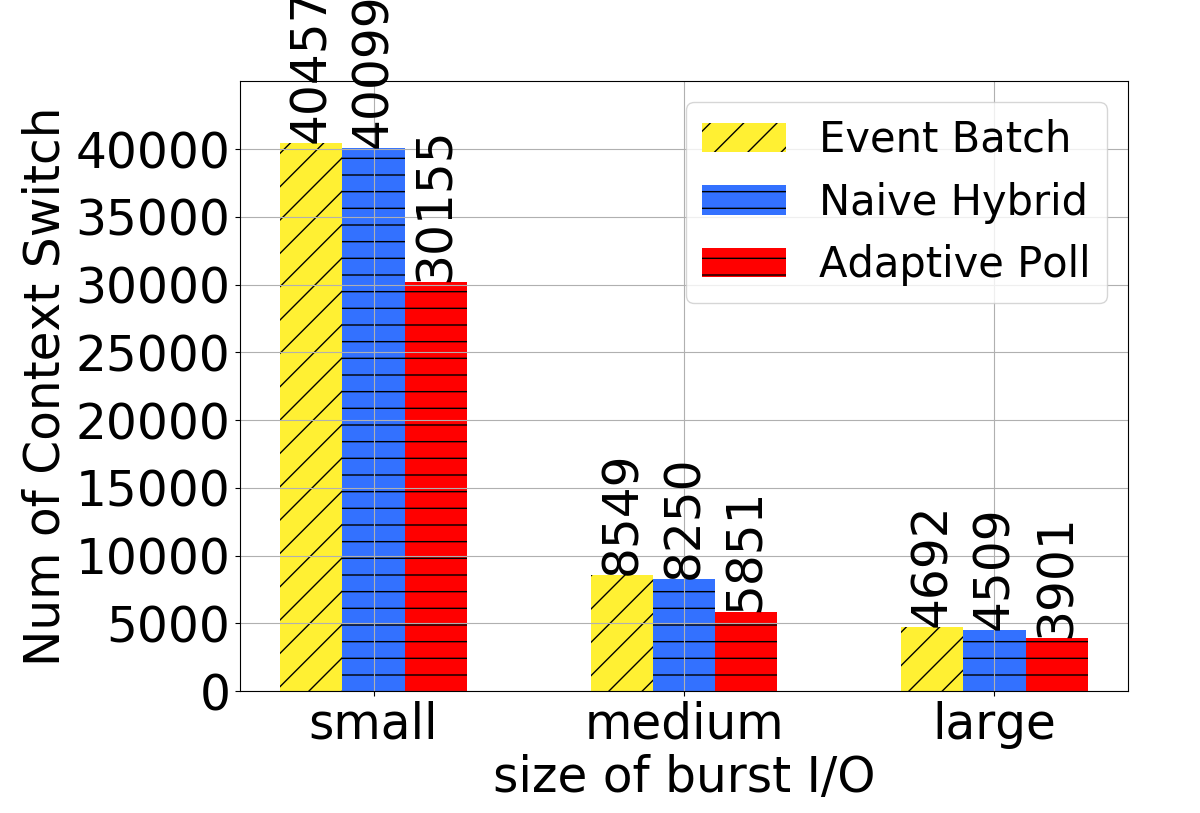}
 \caption{Context Switch} \label{microrst:1d}
\end{subfigure}\hfill
\bigskip
 \caption{ \textbf{Comparison with existing best practices.} Both Event-Batch and Naive Hybrid approach do not adapt well on various big data and ML workload.}
 \label{microrst}
\end{figure}

%---------------------------------------------
\bigskip
\subsection{Poor adaptation of WC handling on various workload}
\label{limitWC}
Existing approaches have limitations in terms of poor adaptation on various workload in big data and ML workload, parallelism and scalability. We compare existing best practices and discuss limitations here. We build micro benchmark to measure the performance of existing best practices. In big data and ML workload, we often observe three I/O patterns: a high intense workload where I/O is issued continuously with a very short interval, a medium intense workload where a pack of I/O is issued with a small interval and a low intense workload where an I/O is issued with a large interval. We observe these workload patterns from logistic regression in our evaluation(\cref{evaluation}), and simulate three types of workload above mentioned(See Fig.\ref{microrst}) by generating a different size of cluster of dense I/O requests with different intervals. We call them small, medium and large in Fig.\ref{microrst}. Workload small is I/O pattern with one request sequentially. It waits until before request is done like other microbenchmark\cite{mellanoxbench}. We then find the smallest size that busy polling can have maximum throughput. For example, if busy polling reaches to highest possible throughput at size K clustered I/Os, it doesn't get higher throughput even if we use larger size of clustered I/O than K. We set this size as workload large. Lastly, for medium workload, we set the median of cluster size between small and large cases. Experiments(\cref{experiments}) and evaluation(\cref{evaluation}) with real world workload are in later sections.

%\textbf{Busy polling}\cite{HERD16,FASST,FARM} is used to maximize the performance at a cost of CPU overhead. Busy polling helps to reduce latency of polling and improve performance but it burns CPU even when I/O is in idle. When the number of remote connections increases, CPU overhead also increases linearly.

%\textbf{Event-triggered mode}\cite{Infiniswap} has no CPU overhead but longer latency compared to busy polling due to context switch and interrupt delay. Unlike Busy polling, Event-triggered mode polls in interrupt context. One interrupt context is required to process one WC item.
We now describe the limitations in existing approach.
\textbf{Event Batch}\cite{nbdx} returns to event mode after a batch is processed regardless of remaining jobs in the queue. It is triggered by next round of interrupt to process the rest of jobs in the queue. Therefore, it shows the highest number of interrupts and context switching compared to others(Fig.\ref{microrst:1c},\ref{microrst:1d})

\textbf{Hybrid(event + busy) polling}\cite{X-RDMA} doesn't adapt well with various workload pattern. Hybrid approach shows better result than Event Batch in high intense workload but its performance is suboptimal in medium and low intense workload because it immediately returns to event mode without catching packs of I/O arriving with some intervals(Fig.\ref{microrst:1a}). This is because it returns to event mode immediately when it fails to poll. When a next pack of I/O arrives shortly, interrupt is triggered again. In this way, there are unnecessary massive number of interrupts and context switching occurred in small and medium cases and it causes suboptimal performance. As we see in Fig.\ref{microrst:1c},\ref{microrst:1d}, it shows comparable number of interrupt and context switching to Event-batch which depends on event triggered mode.

\textbf{Adaptive Polling.}\cite{LITE} We introduce Adaptive Polling shows that it adapts well to various pattern of workload. It shows best performance over other two similar approaches in all workload pattern in Fig.\ref{microrst:1a}). Compared to others, it reduces unnecessary interrupts and context switching(Fig.\ref{microrst:1c},\ref{microrst:1d}). We discuss design of Adaptive Polling(\cref{adaptivepolling}) and analysis with experiments(\cref{experiments}) in later sections.

%-------------------------------------------------------------------------------
\begin{figure}[ht!]
\centering
\includegraphics[width=80mm]{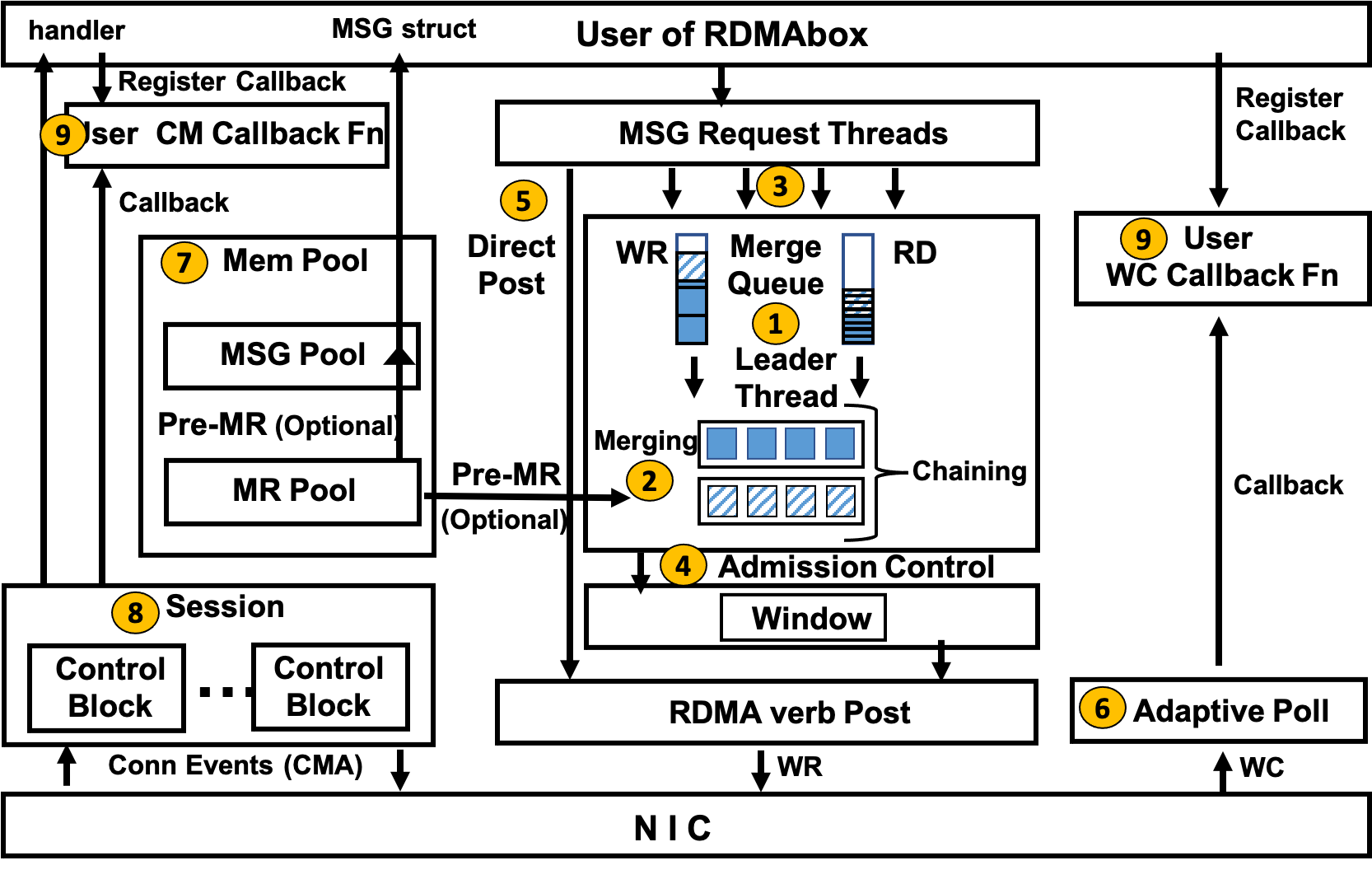}
\caption{ In \textbf{RDMAbox architecture overview}}
\label{architecture}
\end{figure}

\bigskip
\section{RDMAbox architecture}
\label{designoverview}
We provide an overview of our optimization techniques and architecture of RDMAbox library in this section. 
%We package them into a simple I/O abstraction. Application and system developers can  utilize network level abstraction to build their own RDMA-based remote memory system with RDMAbox.
Note that the design principle of RDMAbox is same in both kernel and user space. They slightly differ in implementation or libraries to use.
For more information about low level APIs, we share our open source project on GitHub.
%We provide an overview of our optimization techniques as a network level abstraction in both kernel and user space in this section.

%-----------------------------------
\subsection{Merging and Chaining}
\label{messagebatching}
\noindent

\textbf{Fig.\ref{architecture}-\circled{1} RDMA I/O Merge queue.}
RDMAbox introduces a single merge queue for each write and read for cross-cpu/thread I/O merging at RDMA sending level under two major rules. First, it merges adjacent requests that have the same destination(Fig.\ref{architecture}-\circled{2}). This helps to reduce the total number of RDMA I/Os to NIC. Second, it batches differently based on the load on the merge queue. It doesn't enforce merging when the workload is not high.

\bigskip
\textbf{Fig.\ref{architecture}-\circled{2} Merging and Chaining.}
RDMAbox introduces merging-on-MR to reduce the number of RDMA I/O and chaining with doorbell batch to save the bandwidth of PCIe~\cite{HERD16}. Both techniques work together as a hybrid approach to reduce I/O cost of RDMA. Merging-on-MR first tries to merge adjacent data requests that are on virtually contiguous on the remote memory. By merging multiple requests into one RDMA request, it can save bandwidth of PCIe between NIC and CPU by reducing the number of handshake between CPU and NIC. For instance, if N requests are merged into 1 request, then we can save the space in WQE cache of NIC and reduce N MMIO to 1 MMIO to NIC. If neighbor requests are not adjacent, then it can be chained with doorbell batch to save some of bandwidth in PCIe.

\begin{algorithm}[t]
\small
\caption{reqMSG()}
\begin{algorithmic} 
\STATE \textbf{Input} : $req\_msg$
    \STATE $ enqueue(req\_msg)$
    \STATE $ mergeAndChain()$
    \STATE $ return$
\end{algorithmic}
\label{reqmsg}
\end{algorithm}

\begin{algorithm}[t]
\small
\caption{mergeAndChain()}
\label{alg:workloadawarebatch}
\begin{algorithmic} 
\STATE MAX\_CHAINING\_SIZE : max size of request chaining
\STATE $merge\_cnt \leftarrow 0 $
\bigskip
\STATE $ trafficPacerBlocking()$    
\WHILE{merge\_cnt $<$ MAX\_CHAINING\_SIZE}
    \STATE $req\_msg \leftarrow dequeue() $
    \IF{$req\_msg$ is null}
      \STATE return
    \ENDIF
    \bigskip 
    \STATE $ mergeChecking()$ //check address is adjacent
    \bigskip
    \IF{req\_msg is adjacent}
      \STATE $ reqMerging\_dynMR() or\ reqMerging\_preMR()$
      \STATE $continue$
    \ELSE
      \STATE $ reqChaining()$
      \STATE $continue$
     \ENDIF
     \bigskip
    \STATE $RDMA\_verb\_post() $
   \STATE $ merge\_cnt \gets merge\_cnt + 1$
\ENDWHILE
\end{algorithmic}
\label{mergeandchain}
\end{algorithm}

\begin{figure}[ht!]
\centering
\includegraphics[width=80mm]{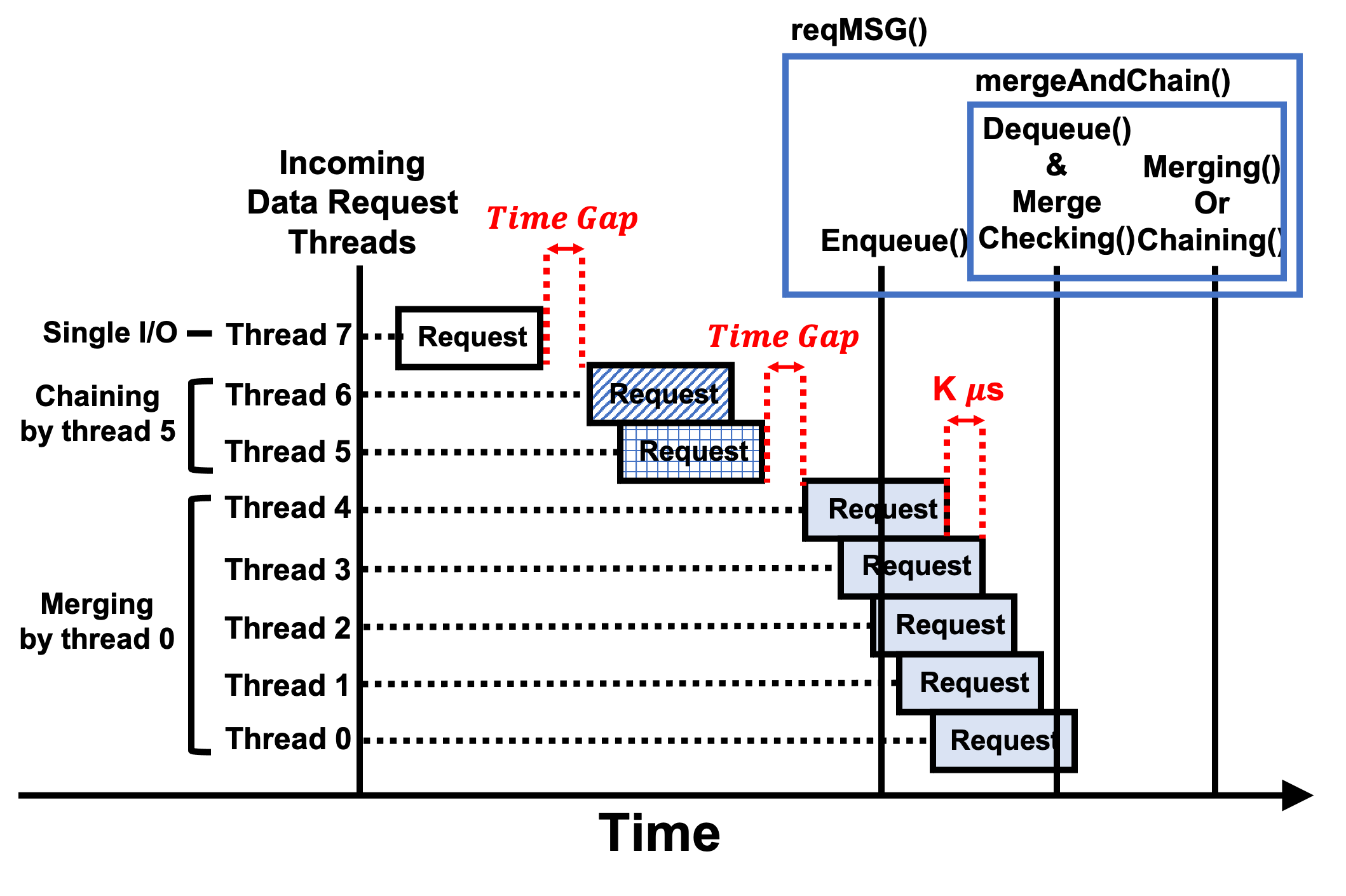}
\caption{ \textbf{RDMAbox merge and chain threads avoid uphill battle by having the earliest thread to process batching and others return.} When thread0 reaches merge checking function, threads[1-4] have already enqueued data requests into the merge queue but not reached merge function yet. Then it is likely that thread0 takes data requests from the queue and makes a batch. Threads[1-4] will return and bring next data request. If thread0 takes too much time on merging, then it is possible that thread 2 or 3 can take the rest of the requests in the first group and do batching instead of waiting. If timing of arrival is close but not adjacent(like thread 5 and 6), it is likely to be batched as a doorbell batch. If request arrives alone(thread7), it will be sent immediately as a single I/O.}
\label{eventtiming}
\end{figure}

\bigskip
\textbf{Fig.\ref{architecture}-\circled{3} Load-aware merging and chaining.}
RDMAbox provides cross-thread I/O merging and chaining but never enforce them and avoids additional latency. The earliest arriving thread checks the merge queue first. If there are more than one request that can be merged, the earliest arriving thread merges data requests. Then, the later arriving thread(s) that originally brought the request(s) into the  merge queue just return(s) since jobs are already taken by the earliest thread. If a request arrives alone later, then it is posted as a single RDMA I/O immediately. See Fig.\ref{eventtiming} for an example scenario. Instead of enforcing merging, it allows merging to happen only when the merge queue is stacked up by many concurrent requests due to heavy workload. Otherwise, each thread will post a single RDMA I/O to send its own request. In this way, merging and chaining can merge requests across threads and yet it does not hurt parallelism. RDMAbox also does not hide vanilla verb behind RDMAbox. Users still can use vanilla verb as well as RDMAbox(Fig.\ref{architecture}-\circled{5}).

%\begin{figure}[ht!]
%\centering
%\includegraphics[width=80mm]{workload_aware_batch.png}
%\caption{ In \textbf{Load-aware Batching across threads}, each data I/O thread enqueues, dequeues and merge-checks. Only a few threads do batching across threads or posting single I/O according to I/O load in the merge queue instead of enforcing cross-thread merging. }
%\label{batchdesign}
%\end{figure}

\bigskip
\textbf{Comparison with existing doorbell batching.}
%\label{doorbell}
Doorbell batching does not reduce the total number of I/O to NIC. Doorbell batching connects multiple WRs with linked list and posts the first WR with MMIO. Then only the first WR is inserted into NIC through MMIO. The rest of chained WRs remains in memory at the moment and are read by DMA-read from NIC. Posting N WRs with Doorbell batch requires one MMIO and N-1 DMA reads. This way saves a bit of bandwidth but does not reduce the total RDMA I/Os(WQEs) to NIC. For instance, in Fig.\ref{eventtiming}, if only doorbell batching is used, the first group will be batched as a doorbell batch but its number of RDMA I/O is as same as that of single I/O case. We provide performance measurements to compare approaches with real world workload in \cref{workloadbatching}

\begin{figure}[ht!]
\begin{subfigure}{0.23\textwidth}
 \centering
 \includegraphics[width=1\linewidth]{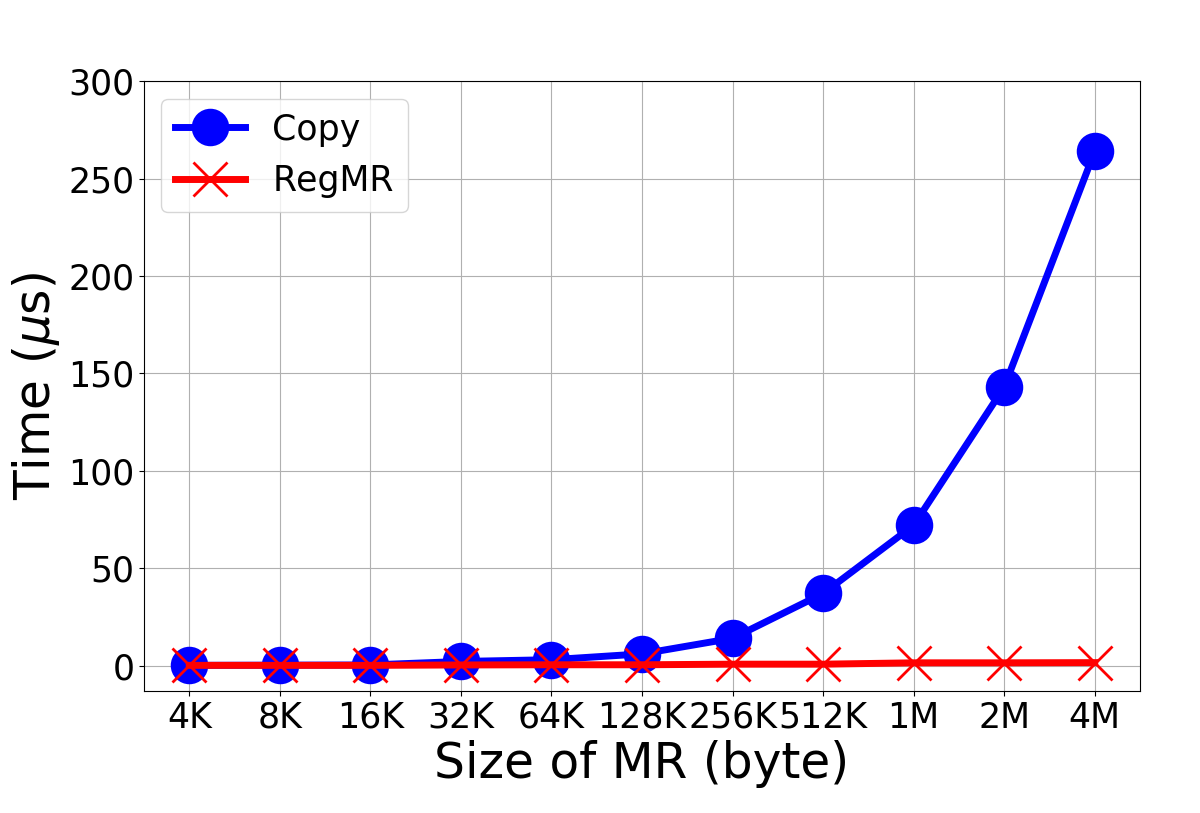}
  \caption{Kernel space} \label{copyvsmrerg:1a}
\end{subfigure}\hfill
\begin{subfigure}{0.24\textwidth}
 \centering
 \includegraphics[width=1\linewidth]{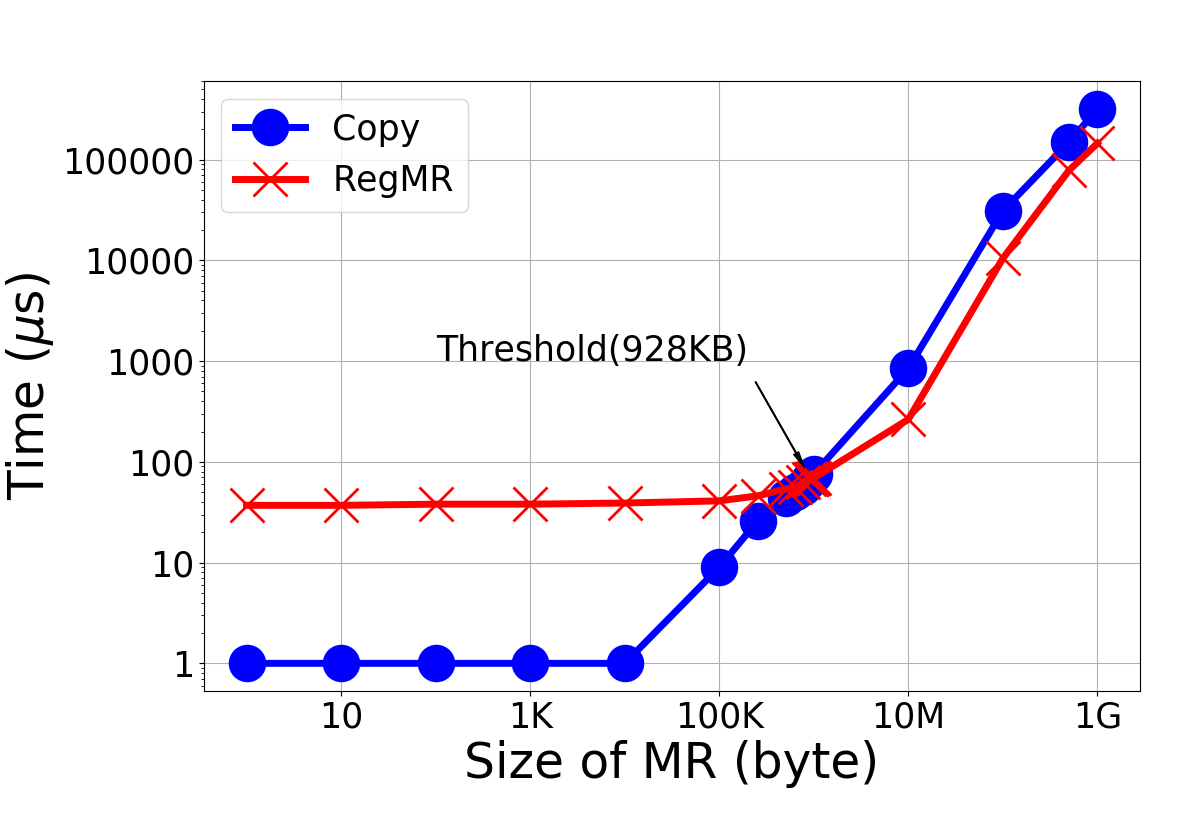}
  \caption{User space} \label{copyvsmrerg:1b}
\end{subfigure}
\bigskip
\caption{\textbf{MR registration vs Memcpy with resident page in kernel space and user space.} Note that scale is different. In kernel space, dynMR is favored at all sizes. In user space, cost of preMR and dynMR is different based on the threshold. }
\label{copyvsmrerg}
\end{figure}

\bigskip
\textbf{Pre-registered MR vs dynamic MR registration.}
%\label{copyvsmrreg}
On top of merging and chaining, there are two ways to batch on MR. One is with pre-allocated and registered MR(preMR for the rest of the paper) and the other is with dynamic MR registration on data buffer(dynMR for the rest of the paper) with Scatter Gather Entry(SGE). PreMR avoids allocation and MR registration cost but entails memcpy from data buffer to MR. DynMR does not have allocation cost but MR registration cost. Frey et al.(2009)~\cite{philip2009} reported that memcpy is faster than MR registration for small memory region(<256KB) in user space. Since virtual address of each page is used in MR registration in user space, overhead of storing address translation and PTE cache in NIC is larger than the cost of copying to pre-registered MR for small memory region. However, we found that the result is different in kernel space MR registration. Unlike in user space, copy cost dominates at all MR size in kernel space(Fig.\ref{copyvsmrerg:1a}). Since physical address is used to register MR in kernel space and it does not have overhead of address translation and PTE cache on NIC. In user space, threshold that MR registration gets better is 928KB in our measurement(Fig.\ref{copyvsmrerg:1b}). Therefore, we recommend dynMR for all message size in kernel space solution and mixed approach of dynMR and preMR based on the threshold in user space.

%-----------------------------------
\bigskip
\textbf{Fig.\ref{architecture}-\circled{4} RDMA I/O level Admission Control.} \label{pacerdesign}
%Traffic pacer is typically implemented with an extra layer of queue(s).
Queueing I/O traffic when I/O channel is congested is a simple and widely used design in many networking systems. Multi-queue based traffic pacer, however, has synchronization cost and fairness issues~\cite{Carousel,Eiffel}. Our rule is simple. We want a single-queue-based traffic regulator without adding an extra layer of queue. By implementing a regulator on a single RDMA I/O merge queue, RDMAbox avoids an extra layer of queue and overhead from multi-queue design. RDMAbox uses window based in-flight data limiter with page granularity. Window size can be up to an upper-limit of NIC capability. This upper-limit is configurable at initialization time of RDMAbox. Fragmentation size also can be adjustable. In remote paging system example(\cref{experiments}), it is equal to block I/O size.

\textbf{Benefit} of this design is to take an advantage out of behavior of waiting in a queue. When the traffic regulator blocks I/O traffic, requests are simply waiting in a queue doing nothing, which is usually a necessary waste of time. Since our design of regulator is implemented on RDMA I/O merge queue, it actually has an extra chance to merge neighbor requests while pacing the traffic. merging and chaining can combine requests while queueing and reduce the number of RDMA I/O and handshake cost, in turn, it improves the performance. RDMAbox also provides a software hook to implement customized admission control policy. For now, we use static window size for traffic regulator in our prototype since our goal in this paper is not to build complete traffic shaping or network congestion control algorithm. Our simple implementation works well and serves the purpose(Fig.\ref{flowcontrol}). Further, by providing the software hook to implement network congestion control solutions~\cite{TIMELY,HPCC}, RDMAbox can extend its capability by implementing such existing software solutions.

%-----------------------------------
\subsection{Adaptive Polling}
\label{adaptivepolling}
%\bigskip
%\noindent
\textbf{Fig.\ref{architecture}-\circled{6} Adaptive Polling}
We propose Adaptive polling and discuss the design decisions in this section. Adaptive polling has two main advantages compared to previous approaches. 

First, Adaptive polling has optimized throughput by combining busy polling and event mode. Unlike previous naive hybrid, Adaptive polling has a hook to catch packs of events that are close each other but with a small time gap. This avoids unnecessary slow interrupt handling and overhead of context switching when processing various pattern of I/O workload.

Second, it has better parallelism than SCQ. In Adaptive polling, RDMAbox allows to have N CQs for N RDMA channels without requiring CPU overhead. This gives better parallelism than serialized single SCQ and one busy polling thread to process it(N CQs vs 1 CQ). It also shows good scalability because it goes back to event mode when there is no events in the queue.

Fig.\ref{microrst} shows that, in all workload pattern, Adaptive polling shows the best performance because it avoids unnecessary interrupts and context switching.

\begin{algorithm}[t]
\small
\caption{Adaptive polling}
\begin{algorithmic} 
\STATE \textbf{Input} : CQ(Completion Queue) 
\STATE MAX\_POLL\_WC : Pre-defined tuning parameter
\STATE MAX\_RETRY : Pre-defined tuning parameter
\STATE initialize $wc\_arr[]$
\STATE $retry \leftarrow 0 $
\WHILE{$true$}
\STATE $num\_wc \leftarrow 0 $

\FOR{num\_wc $\leqslant$ MAX\_POLL\_WC}
  \STATE $rst \leftarrow poll\_cq(CQ , wc\_arr[MAX\_POLL\_WC]) $
    \IF{rst $=$ 0}
      \STATE break
    \ENDIF   
  \STATE $ num\_wc \gets num\_wc + 1$
\ENDFOR

\bigskip
  \STATE $ i \gets 0$
\FOR{ $i < num\_wc$}
  \STATE $ \textbf{handle}( wc\_arr[i] )$
  \STATE $ i \gets i + 1$
\ENDFOR

\bigskip
    \IF{num\_wc = 0}
        \IF{retry $\geq$ MAX\_RETRY}
          \STATE request\_to\_notify\_cq
          \STATE return
    \ENDIF   
      \STATE $ retry \gets retry + 1$
    \ENDIF   
  
\bigskip  
\ENDWHILE
\end{algorithmic}
\label{ap_algo}
\end{algorithm}

\bigskip
\textbf{Design of a hook to catch intermittent and burst load.}
In Adaptive Polling, it adapts well with different workload pattern(\cref{limitWC}). First, with intermittent workload, if the time gap between two events is larger than iteration time of MAX\_RETRY, it behaves as event mode and incur less CPU overhead. Second, with burst and heavy workload, it does not return if burst workload is enough high for Adaptive Polling's hook to detect following events during the iteration(MAX\_RETRY). See Algorithm \ref{ap_algo}. In our benchmark(\cref{limitWC}), we use MAX\_RETRY = 120. If we set large MAX\_RETRY, it works like busy polling. Likewise, if we set small MAX\_RETRY, it works like Event mode. 

\bigskip
\subsection{API design}
\label{apidesign}
RDMAbox utilizes internal mempool for caching message and/or Pre-MR(Fig.\ref{architecture}-\circled{7}) to avoid allocation overhead in the critical path. Session(Fig.\ref{architecture}-\circled{8}) has all necessary information including connection(Control Block) to each peer and internal stateful information. Its handler is provided to user and user can access RDMA components as if they use verb to configure options and parameters in RDMAbox. User also can register handlers for message and CQ event as a callback(Fig.\ref{architecture}-\circled{9}). The rest of the detailed API is in our open source project in GitHub.

%-------------------------------------------------------------------------------
\bigskip
\section{Experiments and Discussions}
\label{experiments}

%RDMAbox also provides a node level abstraction that provides user-transparent remote memory access to user application. RDMAbox implements virtual block device that is connected to remote nodes to achieve this and manages remote resources, data distribution and tracking, and connections. For instance, user can mount RDMAbox block device on a directory and have easy access to remote memory through POSIX file interface(Remote File System) or set it as a swap space for Remote Paging System. One of the main benefit of node level abstraction is that it does not require modification of user application or OS. For remote server daemons to provide and manage remote memory, we follow the design in recent research effort~\cite{valet}.

In this section we evaluate our optimization techniques and provide detailed analysis. Remote paging is useful in data center because it is difficult to predict the workload of memory intensive data center applications\cite{valet}. To reduce the performance gap between local disk and remote memory in remote paging, RDMA in kernel side is essential in design. We build remote paging system with RDMAbox on top of Valet\cite{valet} but without local cache optimization of Valet for fair RDMA performance measurement with real world workload in a data center setting. We mainly present the performance of our approaches here and comparison with other existing approaches will be presented in next section(\cref{evaluation}).

%-----------------------------------
\bigskip
\subsection{Performance of Merging and Chaining}
\label{workloadbatching}

%\bigskip
%\noindent
\textbf{Setup and methodology.}
We use a one-to-one connection to run VoltDB\cite{VoltDB} with YCSB\cite{YCSB}. We use two Facebook workload\cite{Facebookworkload} to create 20GB ETC(Read heavy) and SYS(Write heavy) with Zipfian distribution. We set container limitation to make only 25\% of workload stays in 75\% is swapped out and distributed. We use 128KB block size.

%\bigskip
%\noindent
%\textbf{Batch size impact on performance}
%Batch size(MR size) plays a significant role on performance. We measure performance impact with various batch size(Figure~\ref{batchsize}). We run VoltDB with approach 1 batch optimization with 128KB block I/O. Write batch size starts from 128KB to 4MB. Read batch starts from 4KB to 4MB with 4KB-sized page. Performance increases in both ETC and SYS workload as batch size increases. Although increasing batching size gets better performance, it requires larger MR and increases memory footprint on the host. Therefore, RDMAbox limit the size of total MRs to reduce the memory pressure on the host. Since host node has 32 online cpus in our environment and we set RDMAbox to have 256 depth of submission queues in block layer, RDMAbox logically receives up to 8192(32$\times$256) 128KB-sized requests. By limiting the total size of MR as 1GB(8192$\times$128KB), the best write and read batch size are 512KB and 128KB respectively in our measurement.

%\begin{figure}[!htb]
%\begin{subfigure}{0.23\textwidth}
% \centering
% \includegraphics[width=1.1\linewidth]{batchsize_wr.png}
% \caption{Write Batch} \label{batchsize:1a}
%\end{subfigure}\hfill
%\begin {subfigure}{0.23\textwidth}
% \centering
% \includegraphics[width=1.1\linewidth]{batchsize_rd.png}
% \caption{Read Batch} \label{batchsize:1b}
%\end{subfigure}
% \caption{\textbf{Batch size impact on performance.} Bigger batch size(MR size) gets higher throughput. In this measurement, we increase unit batch size without limiting total size of MR on the host.}
% \label{batchsize}
%\end{figure}

\begin{figure}[!htb]
\begin{subfigure}{0.23\textwidth}
 \centering
 \includegraphics[width=1.1\linewidth]{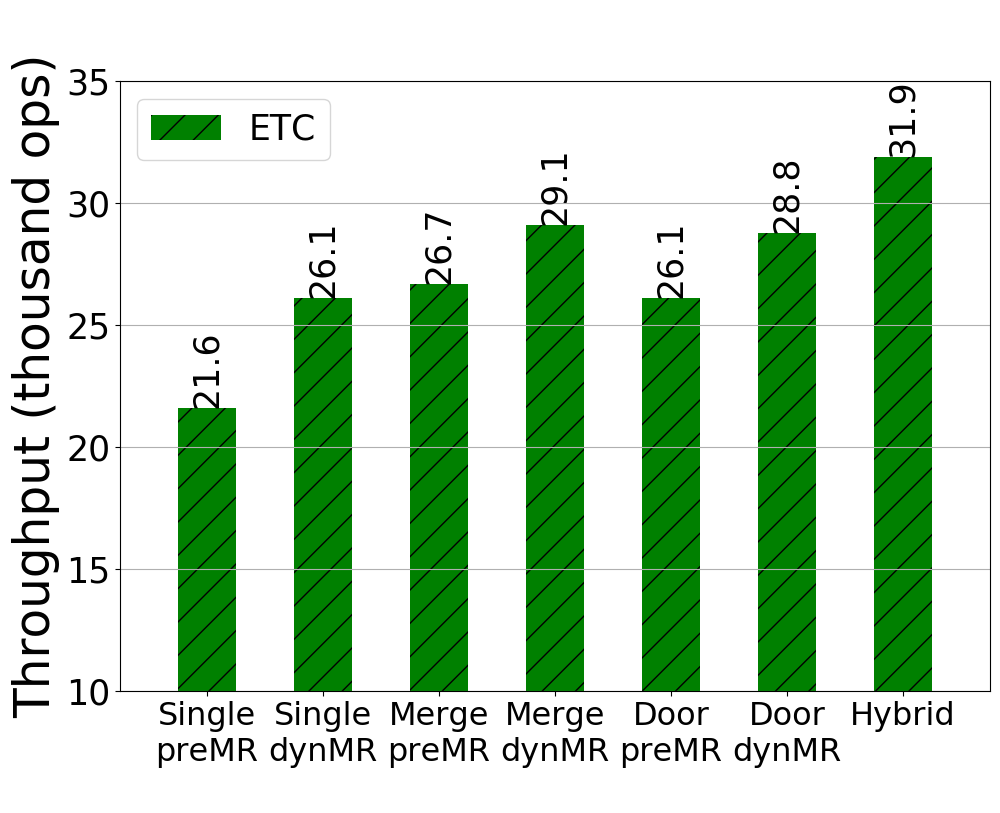}
 \caption{VoltDB ETC} \label{appr_door_etc:1a}
\end{subfigure}\hfill
\begin {subfigure}{0.23\textwidth}
 \centering
 \includegraphics[width=1.1\linewidth]{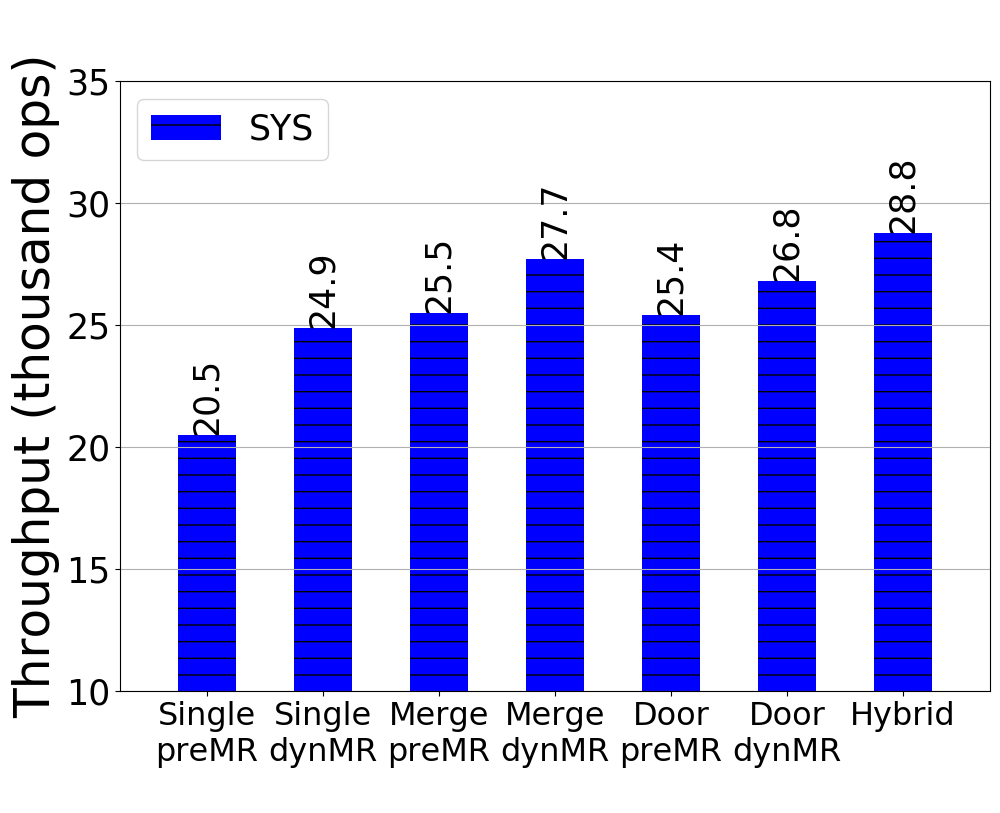}
  \caption{VoltDB SYS} \label{appr_door_sys:1b}
\end{subfigure}
\bigskip
 \caption{\textbf{Comparison of batching approaches}  
 \\ \textbf{Merge} : Merging-on-MR only with preMR or dynMR,  
 \\ \textbf{Door} : Doorbell batching only with preMR or dynMR,
 \\ \textbf{Hybrid} : Merging-on-MR + Chaining with dynMR}
 \label{compappr}
\end{figure}

\bigskip
\textbf{Effectiveness of merging on MR and chaining.}
We compare performance difference among approaches in Fig.\ref{compappr}. In summary, merging-on-MR shows  23.6-24.4\% and 11.2-11.5\%\ improvement over single I/O with preMR and dynMR in both ETC and SYS workload, Hybrid(merging-on-MR + chaining) shows 22.2-47.7\% and 15.7-40.5\% improvement over the single I/O with preMR and dynMR, and 10.8-22.2\% and 7.5-13.4\% over doorbell batch with preMR and dynMR in both ETC and SYS. We also provide our observations here. 

(1) \textbf{Merging-on-MR is better than single I/O}
Merging-on-MR shows better performance than Single I/O(Fig.\ref{compappr}). This is because the number of RDMA I/O is reduced by batching. Table~\ref{rdmaio} reports the total accumulated number of RDMA I/Os measured during running VoltDB ETC workload. It shows that merging-on-MR reduces the number of RDMA I/O well. By having less RDMA I/O, less WQEs toward NIC and less MMIOs by CPUs. N WQEs and N MMIOs can be reduced to one WQE and one MMIO by merging N requests into one. 

(2) \textbf{Merging-on-MR is better than existing doorbell batching.}
Performance of doorbell batch is slightly lower than merging-on-MR because it does not reduce the number of RDMA I/O to NIC. With heavy workload, doorbell batch could still cause a bottleneck in NIC as single I/O does since it doesn't reduce the number of I/Os to NIC.

(3) \textbf{Hybrid take benefits from both batching techniques.}
The bright side of hybrid approach with merging-on-MR and chaining with doorbell batch is that its optimization point is different. Merging-on-MR reduces the number of RDMA I/Os to NIC and doorbell batch reduces bandwidth consumption. The condition where batching happens is also different from each other. Merging-on-MR happens on adjacent requests and doorbell batching can chain non-adjacent requests. Hybrid approach shows the highest performance among others. RDMAbox uses this hybrid approach by default.

%\begin{figure}[ht!]
%\centering
%\includegraphics[width=1.05\linewidth]{comp_appr.png}
%\caption{ \textbf{Comparions of batching approaches} }
%\label{compappr}
%\end{figure}

\begin{table}[t]
\small
\centering
\begin{tabular}{ |p{0.3cm}||p{0.9cm}|p{0.8cm}|p{0.9cm}|p{0.8cm}|p{0.9cm}|p{0.8cm}|  }
 \hline
&Single preMR&Merge preMR&Single dynMR&Merge dynMR&Doorbell dynMR&Hybrid dynMR\\
 \hline
RD&13.2M&\textcolor{red}{11.5M}&13.2M&\textcolor{red}{11M}&\&13.2M&\textcolor{blue}{11M}\\
 \hline
WR&308K&\textcolor{red}{272K}&305K&\textcolor{red}{282K}&\&307K&\textcolor{blue}{280K}\\
 \hline
\end{tabular}
\caption{\textbf{Total number of RDMA I/O to NIC. } We use VoltDB ETC workload. Merging-on-MR shows less RDMA I/O than Single I/O because it is reduced by merging. Hybrid approach has same amount of RDMA I/O to batch since doorbell batching part in the hybrid does not reduce the RDMA I/O. Likewise, doorbell only shows same amount of I/O to single I/O.}
\label{rdmaio}
\end{table}

\begin{figure}[!htb]
\begin{subfigure}{0.22\textwidth}
 \centering
 \includegraphics[width=1.1\linewidth]{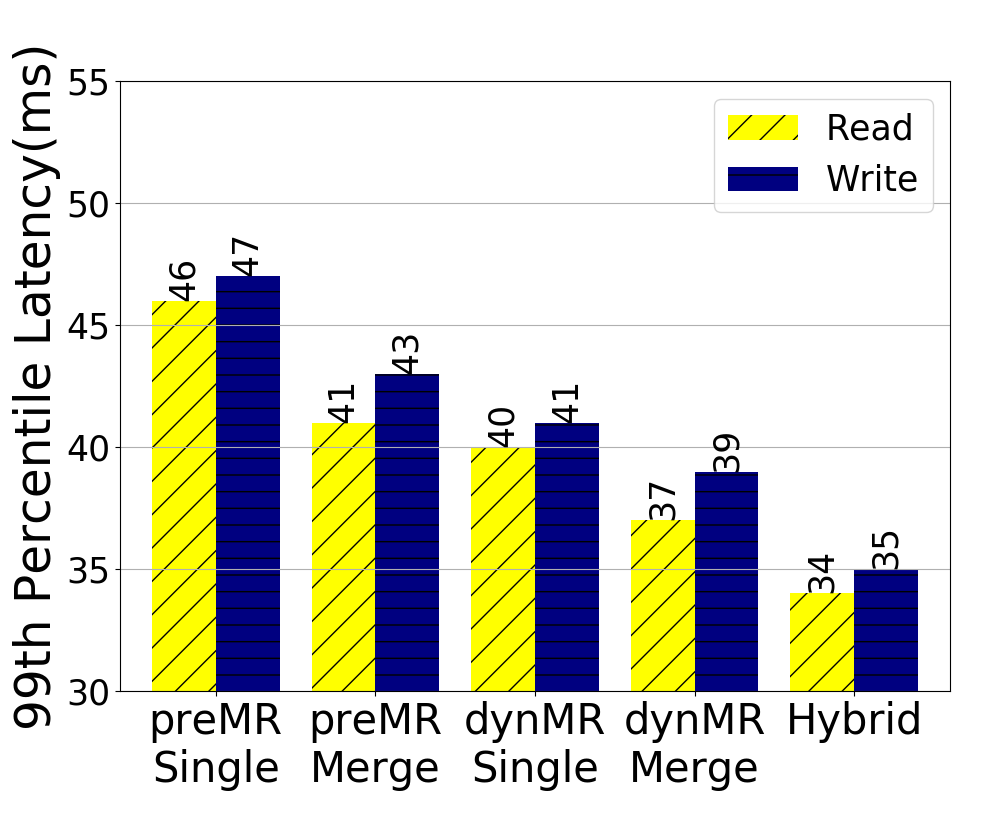}
 \caption{VoltDB ETC} \label{appr_lat:1a}
\end{subfigure}\hfill
\begin {subfigure}{0.22\textwidth}
 \centering
 \includegraphics[width=1.1\linewidth]{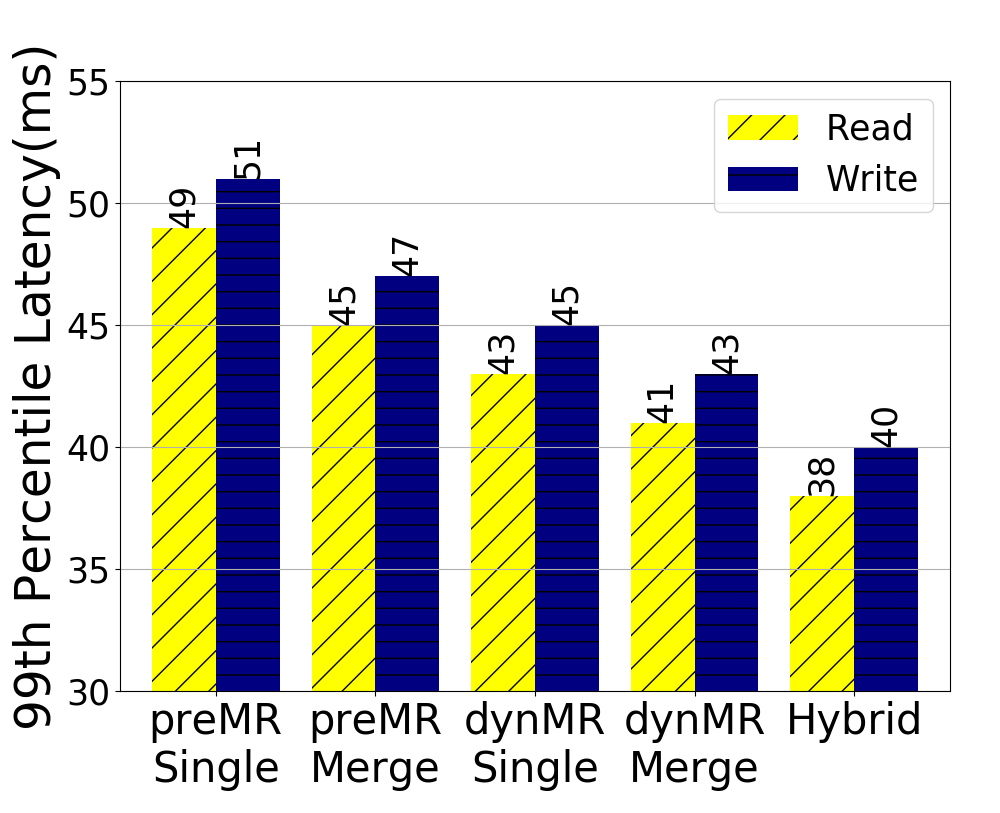}
  \caption{VoltDB SYS} \label{appr_lat:1b}
\end{subfigure}
\bigskip
 \caption{\textbf{Overall application's 99th percentile tail latency with VoltDB ETC and SYS.} Although merging may slightly increase latency per I/O, overall latency is shorter than single I/O that causes more bottleneck in NIC. Hybrid : Merging-on-MR + Chaining. }
 \label{appr_lat}
\end{figure}

\bigskip
\textbf{Latency concerns about Merging and Chaining. }
Merging and Chaining might leave concerns about long tail latency due to cross-thread merging and/or uphill battle among threads. As we mentioned in previous section, RDMAbox doesn't enforce cross-thread merging. If workload is not enough to stack up, RDMAbox sends each of them as a single I/O. A shared structure across threads such as merge queue may cause uphill battle among threads when the workload is intense. However, if the stacking up happens in merge queue, RDMAbox also gets an opportunity to merge the requests. Then, it can reduce handshake cost per single I/O by having one handshake per multiple merged I/Os and, in turn, it reduces overall application latency accumulatively. Without RDMAbox merge queue, stacking up happens inside NIC and there is no chance to merge them while it is stacked up. We measured 99th percentile latency of VoltDB running on remote paging system with various batching approaches. Fig.\ref{appr_lat} shows that our merging-and-chaining does not have negative impact on application latency.

%-----------------------------------
%\subsection{RDMA I/O level Flow Control}

%\textbf{Multi-channel optimization.}
%RDMAbox adopts multi channel optimization to maximize parallelism. The number of channels per remote node is adjustable at initialization time. The number of QPs in the system can be K$\times$N where K is the number of QPs per remote node and N is the number of connected remote nodes. Each channel has its own QP in the dedicated context to avoid false synchronization and limited parallelism due to sharing QPs~\cite{DRTMH,FASST}. In our experiment, 4 channels per remote node setting reaches the best result(See Figure~\ref{multiqp_appr}).

\begin{figure}[!htb]
\begin{subfigure}{0.23\textwidth}
 \centering
 \includegraphics[width=1.1\linewidth]{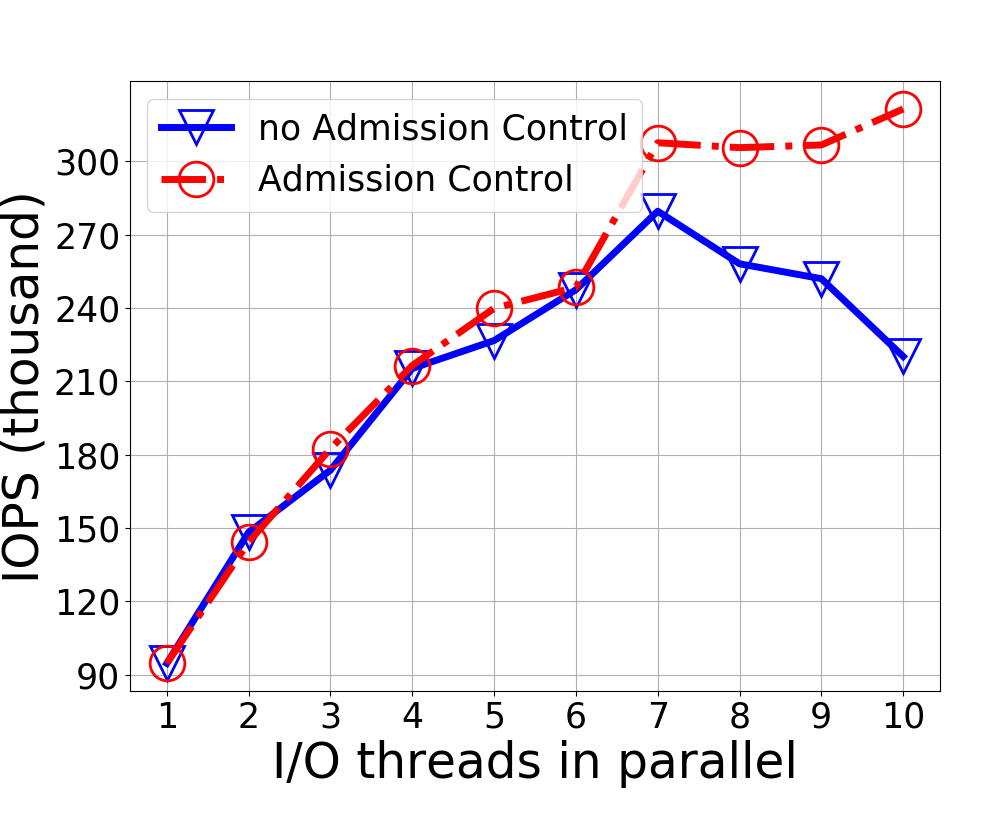}
 \caption{IOPS} \label{flowcontrol:1a}
\end{subfigure}\hfill
\begin {subfigure}{0.23\textwidth}
 \centering
 \includegraphics[width=1.1\linewidth]{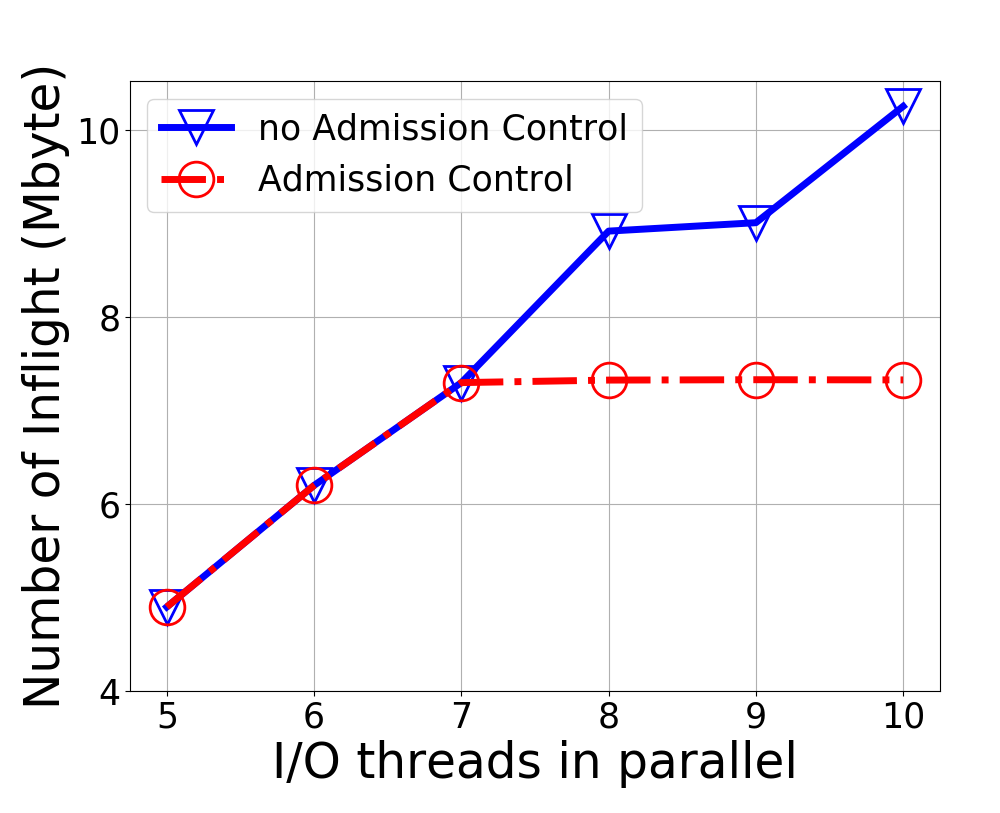}
  \caption{RDMA in-flight byte} \label{flowcontrol:1b}
\end{subfigure}
\bigskip
 \caption{\textbf{With and without RDMA I/O level Admission Control}}
 \label{flowcontrol}
\end{figure}

\textbf{Impact of Admission Control.}
In Fig.\ref{flowcontrol}, we run the same benchmark as we did in Fig.\ref{nic_bottleneck}. Before applying admission control, we try to find the peak performance with multiple QPs. We found that it shows the best performance with 4QPs in the example remote paging system. The peak IOPS is now at 7 I/O threads compared to 4 I/O threads with one QP case in Fig.\ref{nic_bottleneck}. Then, we measure the in-flight byte size at the peak point and use it as a window size of a traffic regulator to show the effectiveness of RDMA I/O level admission control. By pacing in-flight I/O traffic with about 7MB-sized window(Fig.\ref{flowcontrol:1b}), the IOPS increases(Fig.\ref{flowcontrol:1a}). In-flight bytes become stable with the traffic regulator and the highest performance is improved by 29.9\%. 

%-----------------------------------

\begin{figure}[!htb]
\begin{subfigure}{0.23\textwidth}
 \centering
 \includegraphics[width=1.1\linewidth]{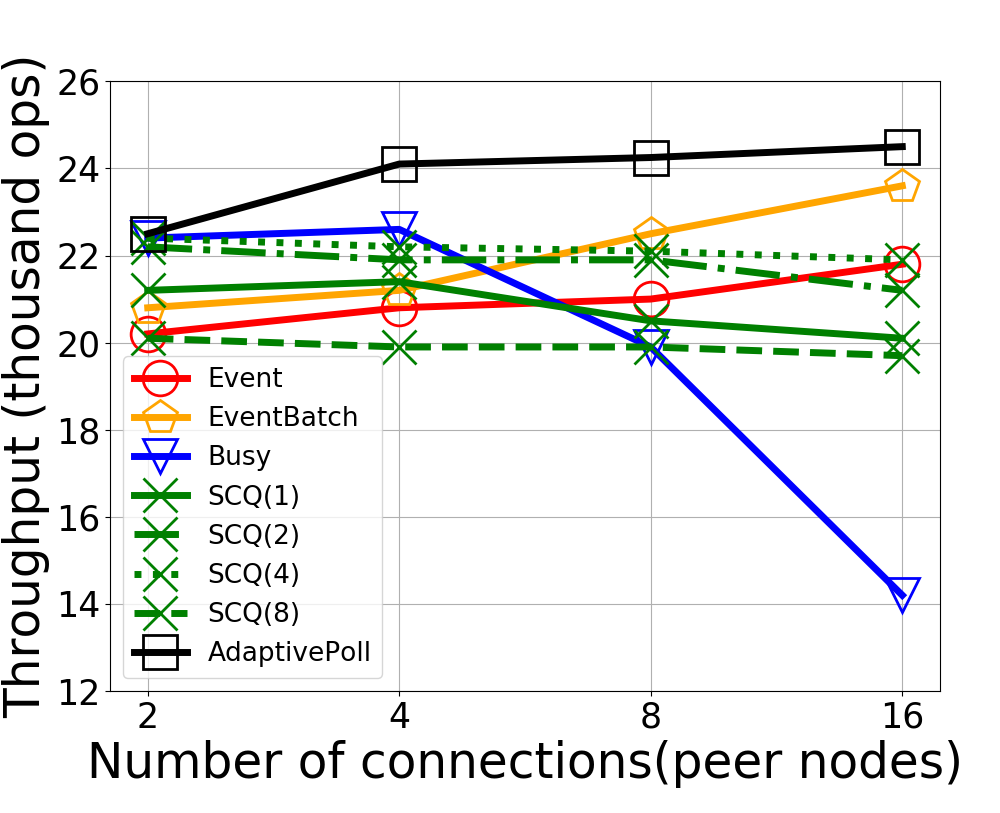}
 \caption{Throughput} \label{pollingexp:1a}
\end{subfigure}\hfill
\begin {subfigure}{0.23\textwidth}
 \centering
 \includegraphics[width=1.1\linewidth]{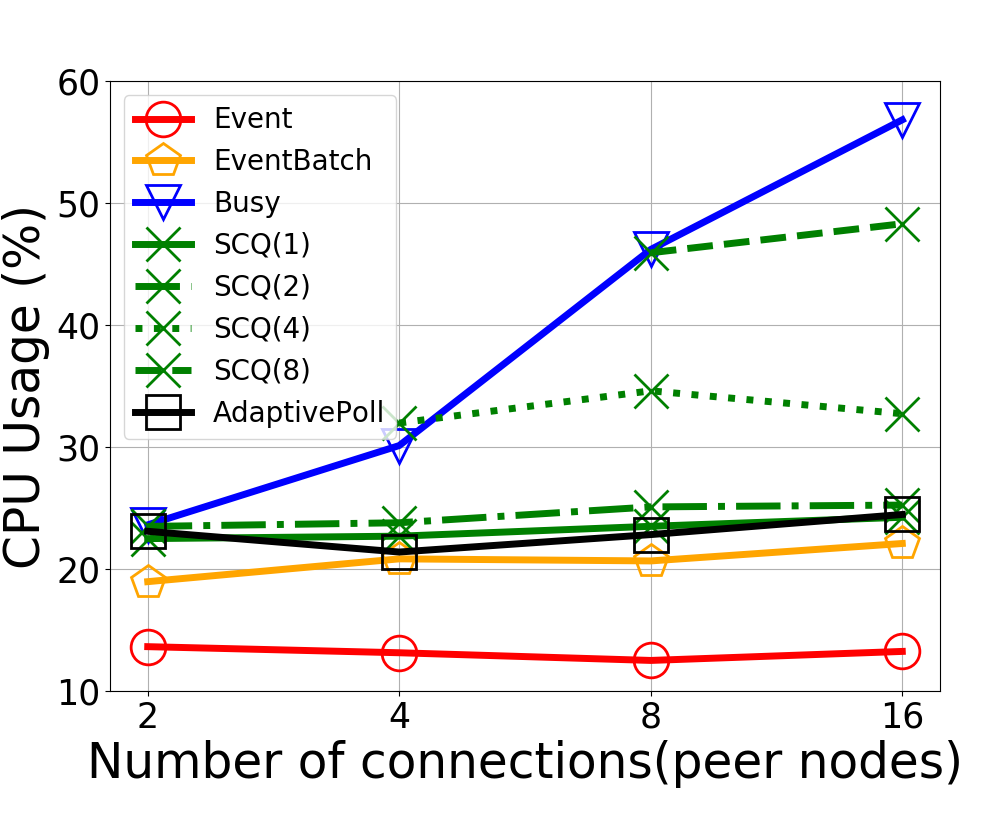}
  \caption{CPU usage} \label{pollingexp:1b}
\end{subfigure}
\bigskip
 \caption{\textbf{Scalability comparison of polling approaches.}
 \\ (N is number of peer nodes.) 
 \\ \textbf{Event} : Event-triggered mode with N CQs. 
 \\ \textbf{Busy} : N busy polling threads with N CQs. 
 \\ \textbf{EventBatch} : Batched Event-triggered mode with N CQs. 
 \\ \textbf{SCQ(M)} : M busy polling thread(s) with M SCQ(s).  
 \\ \textbf{AdaptivePoll} : Adaptive Polling with N CQs.}
 \label{pollingexp}
\end{figure}

\bigskip
\subsection{Scalability of Adaptive Polling}
\label{adaptivepollingexp}
%\bigskip
%\noindent
\textbf{Setup and methodology.} We use one local node and N remote nodes to run VoltDB with SYS workload, which has both intermittent and burst I/O load(\cref{limitWC}). The rest of the setting is the same as in~\cref{workloadbatching} except that we use single I/O with preMR for this experiment. For workload, we use the CPU-intensive VoltDB to evaluate the CPU usage impact of different polling approaches. SYS workload has more write traffic and it causes more CPU activity in OS block layer. We use run-to-completion thread model. We set one channel(QP) per remote node and each channel has one Completion Queue except Shared CQ(SCQ) approach.

\textbf{Poor scalability of Busy polling due to CPU overhead.}
Busy polling shows the best result when it runs with few remote connections(Fig.\ref{pollingexp:1a}). CPU overhead until 4 busy polling threads does not affect the performance. Benefit of busy polling is larger than CPU overhead of 4 busy polling threads. When the number of peer connections increases, throughput severely drops due to CPU overhead(Fig.\ref{pollingexp:1b}). This CPU overhead affects application(VoltDB) performance.

\textbf{Reasonable scalability of Event-triggered mode but suboptimal.}
Event-triggered mode shows higher throughput than Busy polling with many peer connections. Event mode does not have CPU overhead compared to Busy polling(Fig.\ref{pollingexp:1b}) with many peer node connections.

\textbf{Analysis for Shared CQ. }
Shared CQ is shared by multiple threads in a host node. Events from all connections are processed by one busy polling and one Shared CQ, which we denote it as SCQ(1).

SCQ(1) shows worse performance than Event mode with many peer connections. Although Event mode is slow due to interrupt handling and context switch, it has more CQs and more parallel processing context by N CPUs than SCQ(1) with many peer connections(peers$\geq$8). SCQ(1) can be a bottleneck because all WCs are enqueued into this one shared CQ. 

SCQ's parallelism is further limited especially with run-to-completion thread model and/or with replication because run-to-completion increases processing time of each WC and replication increases the number of WCs to process. Fig.\ref{pollingexp} also shows that increasing the number of SCQs doesn't help much.

\textbf{Advantages of Adaptive polling.}
Adaptive Polling shows better parallelism than SCQ. Fig.\ref{pollingexp:1a} shows Adaptive polling gives higher throughput than SCQ in many connections. Adaptive polling also shows better performance and lower CPU overhead than N Busy Polling case. Although Adaptive polling has slightly higher CPU overhead than Event-based approaches, we observe that this CPU overhead does not affect application performance.

\begin{figure*}[!htb]
\begin{subfigure}{0.33\textwidth}
% \centering
 \includegraphics[width=0.5\linewidth]{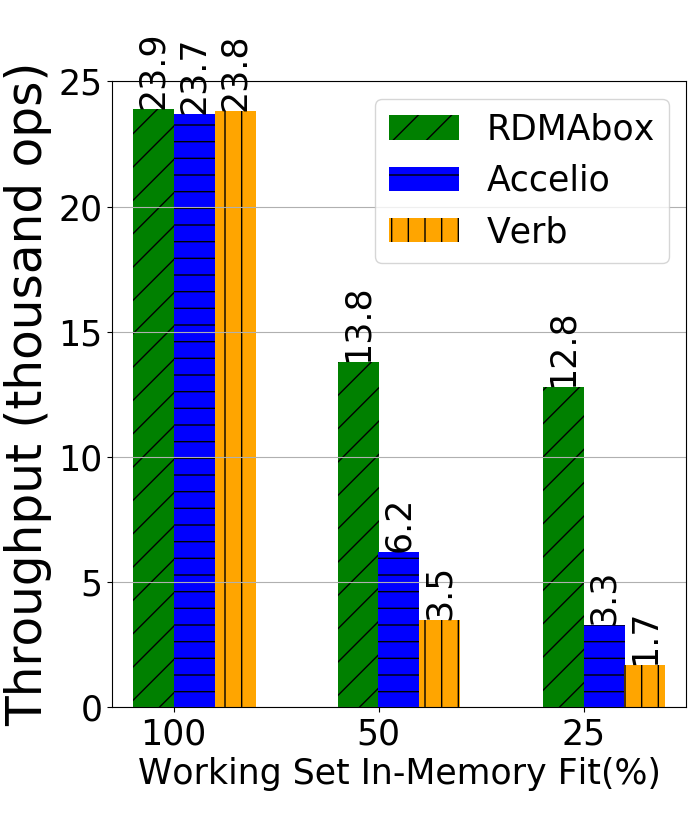}\hfill
 \includegraphics[width=0.5\linewidth]{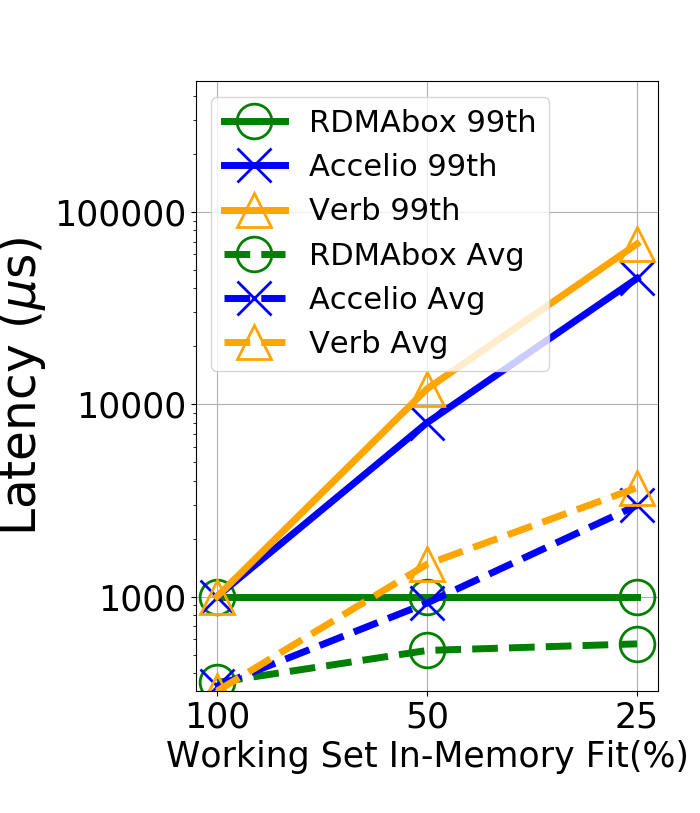}
\caption{MongoDB} \label{eval_bigdata:1a}
\end{subfigure}\hfill
\begin{subfigure}{0.33\textwidth}
% \centering
 \includegraphics[width=0.5\linewidth]{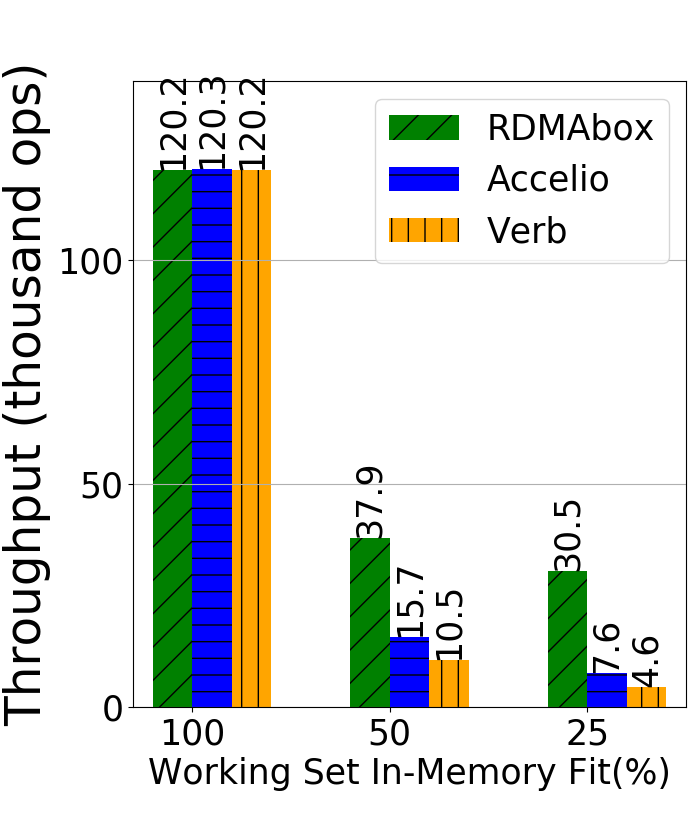}\hfill
 \includegraphics[width=0.5\linewidth]{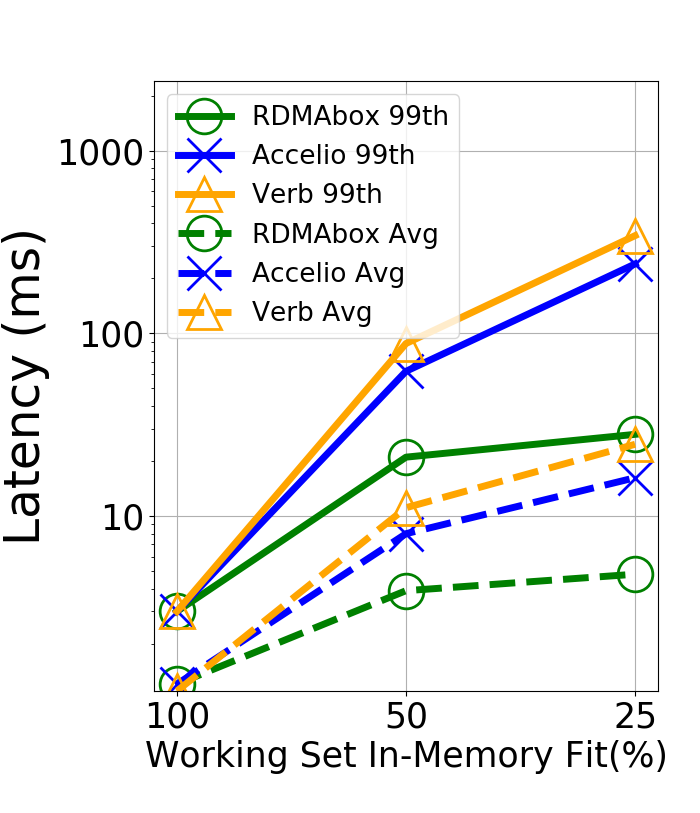}
\caption{VoltDB} \label{eval_bigdata:1b}
\end{subfigure}\hfill
\begin{subfigure}{0.33\textwidth}
% \centering
 \includegraphics[width=0.5\linewidth]{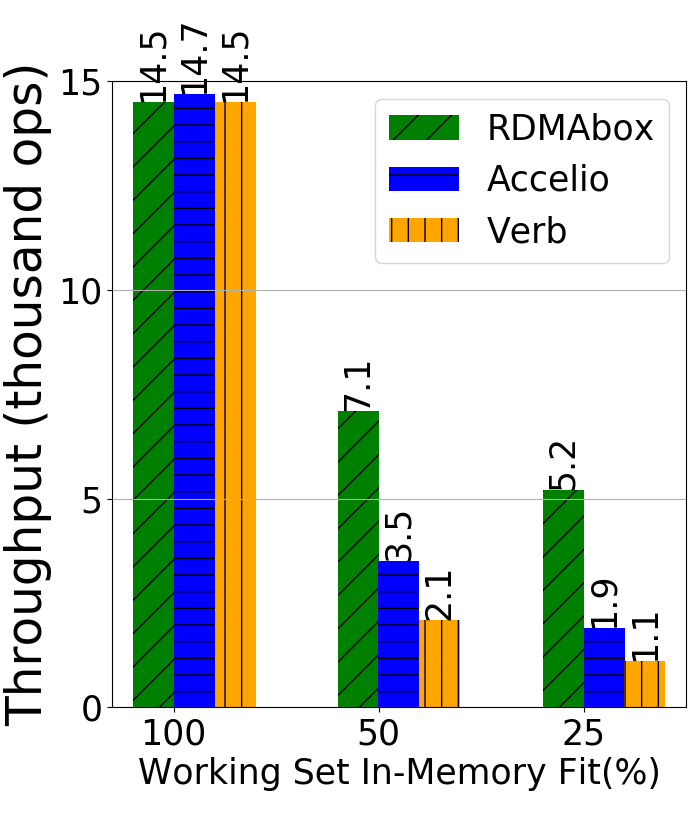}\hfill
 \includegraphics[width=0.5\linewidth]{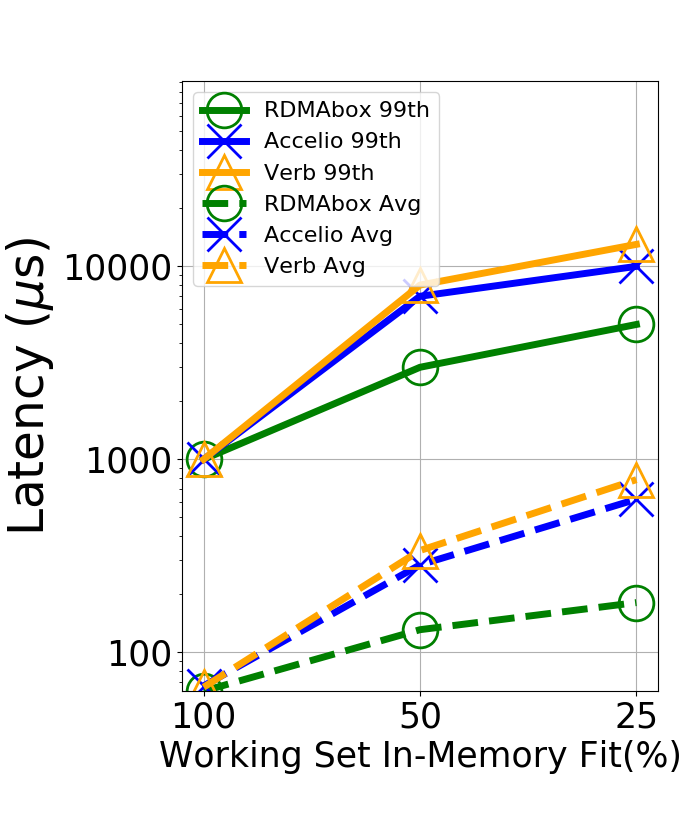}
\caption{Redis} \label{eval_bigdata:1c}
\end{subfigure}
\bigskip
\caption{BigData Applications Comparison with RDMAbox, Accelio and Verb.}
\label{eval_bigdata}
\end{figure*}

\begin{figure*}[!htb]
\begin{subfigure}{0.2\textwidth}
 \centering
 \includegraphics[width=1\linewidth]{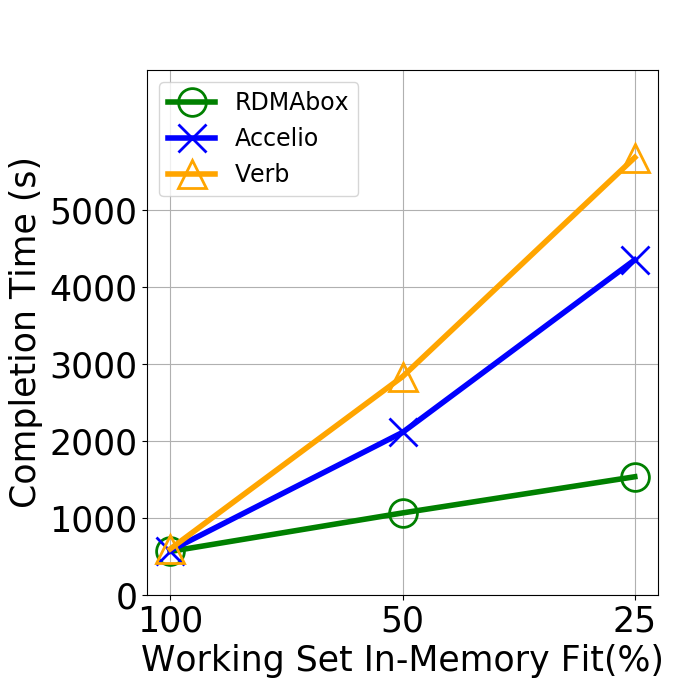}
 \caption{LR} \label{eval_ml:1a}
\end{subfigure}\hfill
\begin{subfigure}{0.2\textwidth}
 \centering
 \includegraphics[width=1\linewidth]{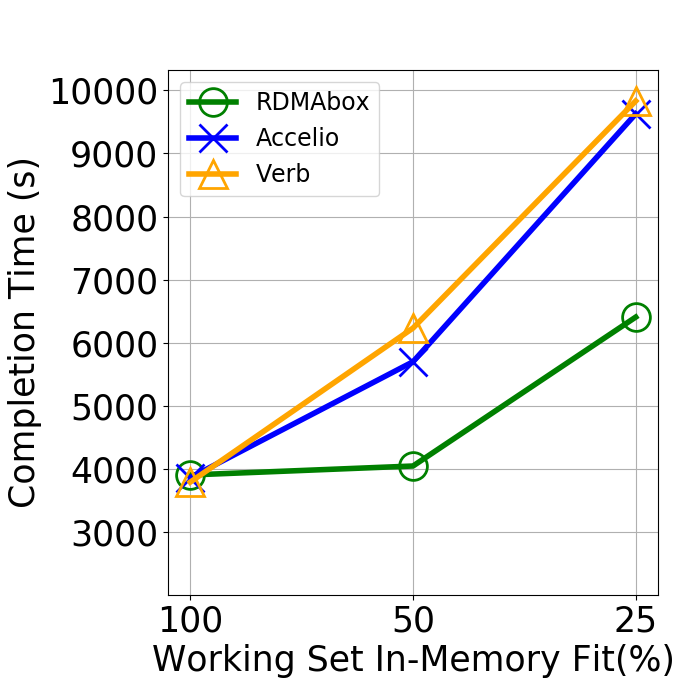}
  \caption{GB} \label{eval_ml:1b}
\end{subfigure}\hfill
\begin{subfigure}{0.2\textwidth}
 \centering
 \includegraphics[width=1\linewidth]{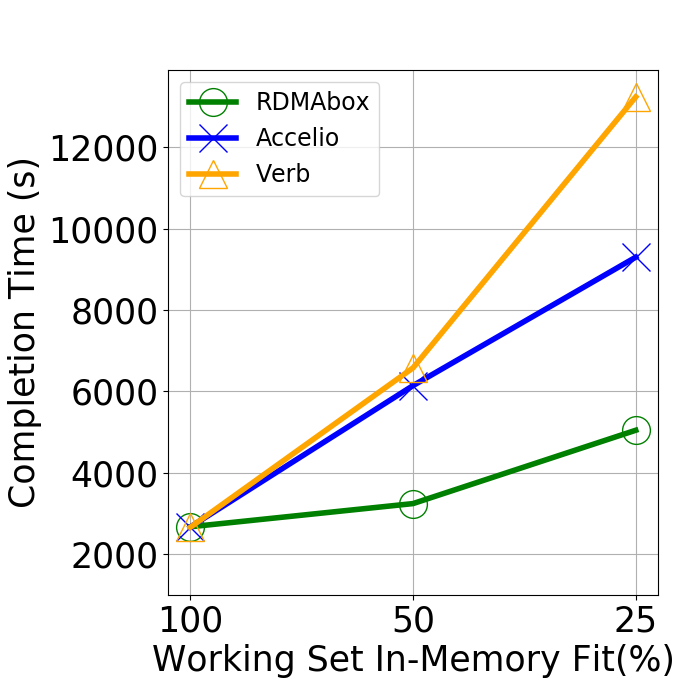}
  \caption{Kmeans} \label{eval_ml:1c}
\end{subfigure}\hfill
\begin{subfigure}{0.2\textwidth}
 \centering
 \includegraphics[width=1\linewidth]{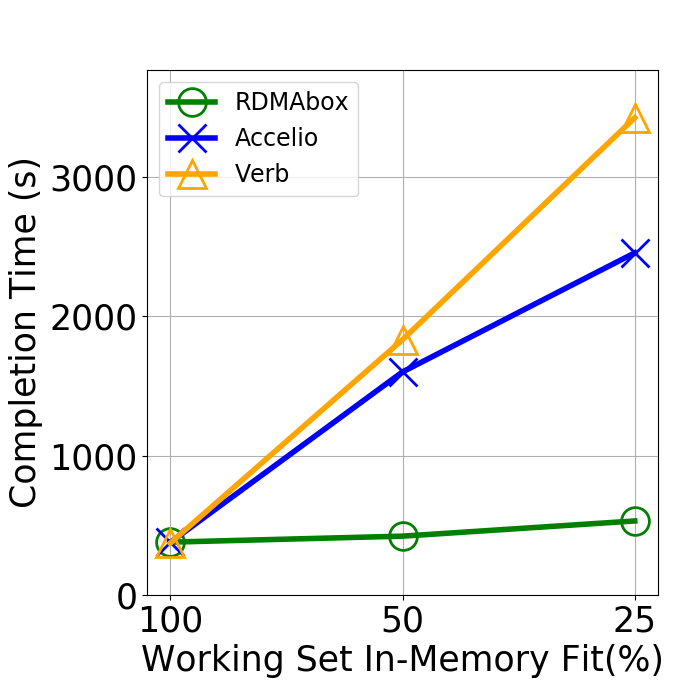}
  \caption{TextRank} \label{eval_ml:1d}
\end{subfigure}\hfill
\bigskip
\caption{ML Applications Comparison with RDMAbox, Accelio and Verb.}
\label{eval_ml}
\end{figure*}

\bigskip
\section{Evaluation with data center applications}
\label{evaluation}
Many efforts have shown that RDMA is beneficial in designing data center applications and systems in both kernel\cite{Accelio, Infiniswap, valet} and user space\cite{orion, glusterfs}. We implement a kernel remote paging system and a user-space network file system to show the performance of RDMAbox compared to existing best practices in both kernel and user space. We report our empirical study and comparison results in this section. %First, we show that the RDMABox based remote paging system outperforms the implementation with Accelio~\cite{Accelio}, with up to 6.48$\times$ throughput improvement and up to 83\% decrease in average tail latency in bigdata workloads, and up to 83\% reduction in completion time in machine learning workloads. Second, the RDMABox based user-space file system on top of FUSE achieves 1.6$\times$ $-$ 6$\times$ throughput improvements over the implementation with existing libraries, represented by Octopus~\cite{octopus}, GlusterFS~\cite{glusterfs} and Accelio~\cite{Accelio}. 

%----------------------------------- 
\subsection{Remote Paging System with RDMAbox}
%\subsubsection{Setup.}
\label{RPCsetup}
\textbf{Setup and methodology.} We evaluate the impact of our optimization on applications using seven memory intensive applications for evaluation. We use three big-data applications: MongoDB\cite{MongoDB}, VoltDB\cite{VoltDB} and Redis\cite{Redis}, and four ML workloads: Logistic Regression\cite{Scikit-learn}, Gradient Boost classification\cite{jia2014caffe}, K-means\cite{PowerGraph} and TextRank\cite{TextRank}. 
All experiments are performed  on Cloudlab\cite{Cloudlab} nodes with Xeon E5-2650v2 processors(32 2.6Ghz virtual cores), 64GB 1.86Ghz DDR3 memory, 1TB SATA 3.5'' rpm hard drives and Mellanox ConnectX-3. We set containers on the host to create swap traffic for our remote paging system example. We first measure peak memory of each application and set container memory limit to create 50\% and 25\% of in-memory working set fit case. Each application runs with total 22GB up to 50GB in-memory working set workload. Finally, we compare RDMAbox with vanilla RDMA verb and the most performant remote paging systems with Accelio, which is kernel space RDMA library.

\bigskip
\subsubsection{\textbf{BigData workload performance}}
%MongoDB is a general purpose, document-based, NoSQL distributed database. VoltDB is a ACID-compliant in-memory transactional database. We set both VoltDB and MongoDB as in-memory-only database. Redis is in-memory distributed caching system through key-value interface. We choose these applications because these are popular ones and have indexing strategies for efficient in-memory computing. This requires more memory for indices as well as dataset and makes workload memory-intensive. These applications show good performance when working set is in memory but it suffers from performance degradation when host node runs out of memory and starts to swap to disk. 

We use YCSB~\cite{YCSB} with Zipfian distribution to create Facebook simulated workload~\cite{Facebookworkload} ETC and SYS. ETC has 95\% read and 5\% write and SYS has 75\% read and 25\% write. We populate these applications with 10 million record first and run 10 million queries to create 15GB to 22GB working set. 

Fig.\ref{eval_bigdata} shows RDMAbox outperforms existing RDMA libarary Accelio up to 3.87$\times$\ in MongoDB throughput, 4.01$\times$\ in VoltDB throughput, and 2.73$\times$\ in Redis throughput. RDMAbox outperformance vanilla RDMA verb up to 7.52$\times$\ in MongoDB throughput, 6.6$\times$\ in VoltDB throughput, and 4.72$\times$\ in Redis throughput. RDMAbox also shows lower latency in both average and 99th tail latency. The average latency is larger up to 5.24$\times$\ in Accelio and 6.47$\times$\ in verb. 99th tail latency is larger up to 45$\times$\ in Accelio and 68$\times$\ in verb compared to RDMAbox. Importantly, the gap between RDMAbox and other systems gets bigger when they rely on more remote memory. 

%-----------------------------------
\bigskip
%\vspace{-9pt}
\subsubsection{\textbf{ML workload performance}}
We evaluate RDMAbox with popular ML workload with actual dataset~\cite{textrankdata, mldata1, mldata2}, which includes 4 to 87 million samples. In Fig.\ref{eval_ml}, completion time of model training in Accelio and verb is higher up to 2.83$\times$\ and 3.69$\times$\ in LogisticRregression, 1.5$\times$\ and 1.53$\times$\ in GradientBoosting Classification, 1.8$\times$\ and 2.62$\times$\ in K-means, and 4.62$\times$\ and 6.44$\times$\ in TextRank than RDMAbox. It shows that memory-hungry workload such as TextRank gets more benefit on RDMAbox. Compute-intensive workloads like K-means and GradientBoosting show less performance gap compared to other workloads.

%\begin{figure*}[!htb]
%\begin{subfigure}{0.33\textwidth}
% \centering
% \includegraphics[width=0.5\linewidth]{scale_thp_mongo.png}\hfill
  %\includegraphics[width=0.5\linewidth]{scale_lat_mongo.png}
%\caption{MongoDB} \label{scale_thp:1b}
%\end{subfigure}
%\begin{subfigure}{0.33\textwidth}
% \centering
% \includegraphics[width=0.5\linewidth]{scale_thp_volt.png}\hfill
  %\includegraphics[width=0.5\linewidth]{scale_lat_volt.png}
%\caption{VoltDB} \label{scale_thp:1a}
%\end{subfigure}
%\begin{subfigure}{0.33\textwidth}
% \centering
% \includegraphics[width=0.5\linewidth]{scale_thp_redis.png}\hfill
  %\includegraphics[width=0.5\linewidth]{scale_lat_redis.png}
%\caption{Redis} \label{scale_thp:1c}
%\end{subfigure}
%\caption{BigData Applications Scalability Comparison between RDMAbox and Infiniswap.}
%\label{scale_thp}
%\end{figure*}

%------------------------------------------------------------------------------
\bigskip
\subsection{Remote File System with RDMAbox}
%To provide user-transparent remote memory access, we build Network File System with RDMAbox node level abstraction. Remote File System is mounted on a directory to provide a user transparent access to remote memory in the cluster. RDMAbox manages RDMA networking including remote resource management. Then, user application can read or write a file on this directory through POSIX file interface. 
%Although key-value store such as HERD~\cite{HERD14} also provides data shipping capability with low-level RDMA optimizations, HERD utilizes UD(Unreliable Datagram) protocol and cannot support larger data size than MTU(up to 4KB in Mellanox ConnectX-3) in one request. It requires data to be fragmented with MTU size and does not favor the design for large size data shipping. 
 Typical ping-pong style benchmark may provide some analysis on latency per operation but it does not represent real world workload because operation is processed one at a time. We build network file system in user space with FUSE support to evaluate RDMAbox user-space library. 
 %Note that, for better performance, this Remote File System can also be implemented in kernel side with RDMAbox kernel-space library that we introduced in section~\cref{experiments}. Since the main advantage of FUSE is portability over performance, many existing file systems provide FUSE interface\cite{fuse file systems}. 
 We compare RDMAbox with existing RDMA middleware such as Accelio\cite{Accelio}, UCX\cite{UCX} and Libfabric\cite{libfabric}, and RDMA-optimized distributed file systems such as Octopus\cite{octopus} and Glusterfs\cite{glusterfs}.

\textbf{Setup and methodology.}
We first compare with existing best practices: Accelio~\cite{Accelio}, UCX\cite{UCX} and Libfabric\cite{libfabric} by implementing FUSE-based network file system. In server, we simply implement in-memory storage that is accessed by one-sided RDMA without server side kernel intervention. Remote regions~\cite{remoteregion} and X-RDMA~\cite{X-RDMA} are also academic and commercial RDMA middleware for remote memory access but RemoteRegion and X-RDMA don't have open source code. Hence, they are not included in our evaluation.

We also compare with RDMA-optimized FUSE file systems such as Octopus~\cite{octopus}, GlusterFS~\cite{glusterfs}. Octopus is designed for a distributed file system with NVMe and RDMA and provides FUSE based API too. In their implementation, they use DRAM to simulate NVMe. Since they also access remote storage with one-sided RDMA, server design doesn't have any meaningful impact on performance. We use FUSE-Octopus, which they provide but it is not reported in the original paper. GlusterFS is also FUSE-based distributed file system that provides flexible and easy deployment of Gluster volume servers on a RDMA-based cluster. We set GlusterFS on ramdisk instead of disk for fair comparison. Note that we only compare raw I/O performance here because metadata management in each system is different. Octopus and GlusterFS have distributed metadata to provide richer distributed metadata management capability. For GlusterFS, it also uses local cache for metadata. For FUSE, we use same version of FUSE client and default options except MAX\_WRITE=128KB.

We run IOzone~\cite{iozone} on a mounted point of FUSE. It opens one test file and issues read and write for total 10GB data. We set up one client and one server to exclude connectivity management from evaluation factors.

\begin{figure}[!htb]
\begin{subfigure}{0.23\textwidth}
 \centering
 \includegraphics[width=1.1\linewidth]{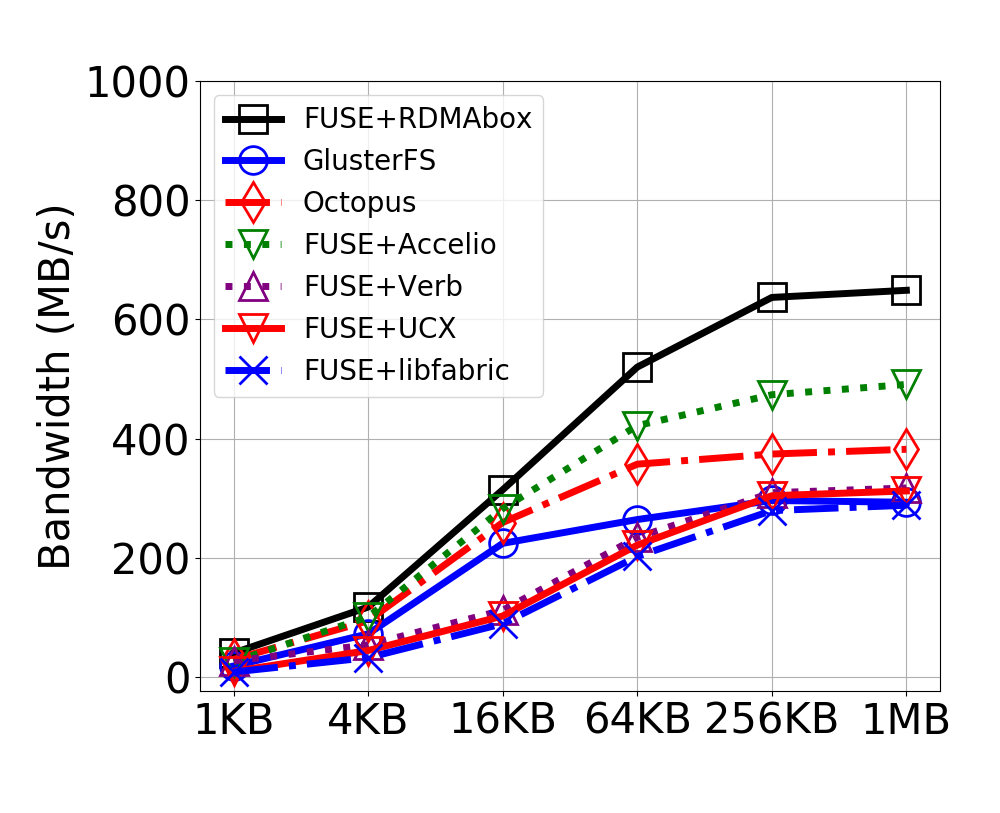}
 \caption{Write} \label{fusethp:1a}
\end{subfigure}\hfill
\begin {subfigure}{0.23\textwidth}
 \centering
 \includegraphics[width=1.1\linewidth]{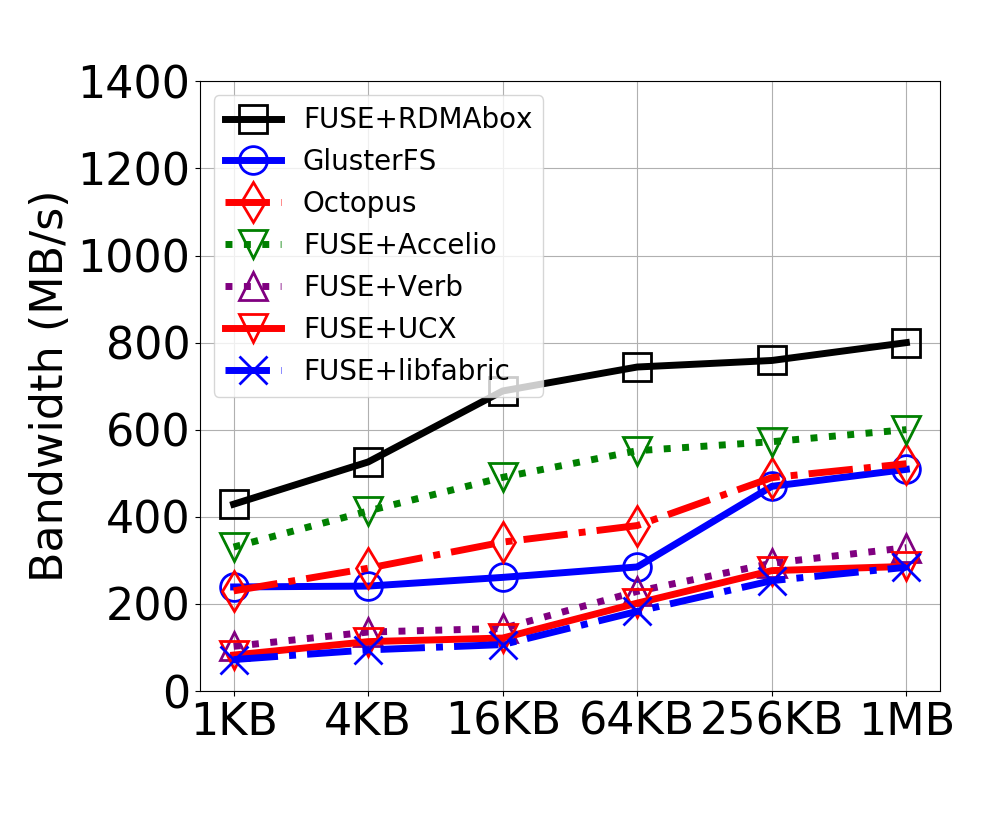}
  \caption{Read} \label{fusethp:1b}
\end{subfigure}
\bigskip
 \caption{Performance comparison of our RDMAbox user space library based remote file system with other existing middleware and remote file systems.}
 \label{fusethp}
\end{figure}

%\subsubsection{Analysis on RDMA optimization}
{\bf Experimental Results.\/} \\
We report the performance comparison results in Fig.\ref{fusethp}. Remote File System with RDMAbox has 1.5$\times$ - 2.8$\times$ higher write bandwidth over Verb, 1.1$\times$ - 1.6$\times$ over Accelio, 2.1$\times$ - 3.1$\times$ over UCX, 2.2$\times$ - 3.7$\times$ over Libfabric, 1.4$\times$ - 1.7$\times$ over Octopus and 1.4$\times$ - 2.2$\times$ over GlusterFS.

In read bandwidth, RDMAbox has 2.4$\times$ - 4.8$\times$ higher over Verb, 1.2$\times$ - 2.6$\times$ over Accelio, 2.7$\times$ - 5.6$\times$ over UCX, 2.8$\times$ - 5.9$\times$ over Libfabric, 1.5$\times$ - 2$\times$ over Octopus and 1.6$\times$ - 2.6$\times$ over GlusterFS.

In RDMAbox, merge-and-chain is used with dynamic switching between preMR and dynMR based on threshold in user space. So, preMR is used on small data size and dynMR on large data size. Adaptive Polling provides optimized performance like busy polling and incurs less CPU overhead with a hook that catches I/O requests with some interval in both sender and receiver side. Admission control and multi QP optimization are also contribution factors. In our experiments, RDMAbox shows the best performance with 4 QPs.

Accelio shows the most comparable performance to RDMAbox because it uses doorbell batch with dynamic registered MR(dynMR) compared to others that use single I/O. However, merge-and-chain shows better performance than doorbell batch(recall Fig.\ref{compappr}), Event-Batch incurs more interrupt handling and context switching with heavy workload(recall Fig.\ref{microrst}), and it also doesn't have RDMA I/O level admission control(recall Fig.\ref{flowcontrol}) like others.

\section{Related Work}
\label{relatedwork}
RDMA optimizations have been proposed across different types of systems mentioned in introduction.

\textbf{Multi QP.}
FaRM shows performance impact of multi QP optimization by varying the number of QPs. Request rate increases as the number of QPs increase, but it decreases as NIC runs out of space for QP cache. Kalia et al. (2016)\cite{HERD16} also utilizes multi QP optimization to engage multiple NIC PU(Processing Unit)s.

\textbf{Selective signaling and inline operation.\/}
Unsignaled verb is used to reduce NIC-initiated DMA (i.e., completion DMA write) and inline RECV in the CQE when the payload is small~\cite{HERD16}. Such efforts reduce DMA from NIC and allow PCIe bandwidth to be used for other operations.
 
\textbf{One sided vs two sided.}
One sided vs two sided verb was actively discussed through key value stores, transaction and RPC systems~\cite{PILAF,FARM,HERD14,FASST,DRTMH}.Pilaf\cite{PILAF} and FaRM\cite{FARM} utilize one-sided RDMA for GET and PUT in key value stores. Unlike previous approach, HERD\cite{HERD14} suggested to use two sided UD SEND with WRITE verb in key value stores. Since UD is scalable and fits for small message size, FaSST\cite{FASST} also utilize two-sided UD SEND for RPC systems. DrTM+H\cite{DRTMH} reported discussions between one-sided and two-sided in transaction systems with many experiments. 

\textbf{Optimizations in RDMA middleware.}
Accelio\cite{Accelio} is RDMA I/O and RPC library for both kernel and user space. It uses Doorbell batch with dynamic registered MR(dynMR), EventBatch, multi QP optimization and two-sided. Doorbell batch with dynMR shows a lot higher performance than single I/O with preMR or dynMR, EventBatch is also good for less CPU overhead with many server node connections. On the other hand, busy polling on Octopus creates CPU overhead on a client node that is connected to many server nodes. UCX\cite{UCX} and Libfabric\cite{libfabric} are only for user space and they focus more on establishing a set of interfaces for implementing multiple programming model libraries and languages for portability than optimization. Remote Region\cite{remoteregion} introduces file interface to remote memory access that is easy and well understandable to programmers.

\textbf{Optimizations in RDMA-based file system.}
Octopus utilizes single I/O with pre-registered MR pool(preMR), busy polling, multi QP optimization and one-sided. GlusterFS uses Single I/O with dynMR, batch polling and two-sided.

%-------------------------------------------------------------------------------
\section{Conclusion}
\label{conclusion}
We have presented RDMAbox with a suite of RDMA optimizations, packaged in easy-to-use kernel and user-space libraries. We motivate the design of RDMABox optimizations with empirical analysis of inherent problems in conventional RDMA I/O operations. To demonstrate the flexibility and effectiveness of RDMAbox, we also implement a kernel remote paging system and a user-space file system. Extensive experiments on big data and machine learning workloads show the effectiveness of RDMABox optimized implementations over existing representative approaches. By reducing the cost of RDMA I/O with our optimizations, remote paging systems with RDMAbox achieve up to 6.48$\times$ higher throughput with a reduction of up to 83\% in 99th percentile tail latency and up to 78\% in average latency in bigdata workloads, and up to 83\% reduction in completion time of model training in machine learning workloads over existing representative solutions. User space file systems based on RDMAbox achieve up to 5.9$\times$ higher throughput over existing representative solutions. %To further improve RDMAbox, low level RDMA driver's support such as a feedback for NIC cache miss will be a valuable information to the admission control management. We leave this exploration as a future work.

%-------------------------------------------------------------------------------
%\subsection*{Acknowledgement}

\end{document}